\documentclass[aos]{imsart}

\RequirePackage{amsthm,amsmath,amsfonts,amssymb}
\RequirePackage[authoryear]{natbib} 
\RequirePackage[colorlinks,citecolor=blue,urlcolor=blue]{hyperref}
\RequirePackage{graphicx}
\RequirePackage{enumitem}

\startlocaldefs

\newtheorem{theorem}{Theorem}[section]
\newtheorem{lemma}[theorem]{Lemma}
\theoremstyle{remark}

\newcommand{\openr}{\hbox{${\rm I\kern-.2em R}$}}
\newcommand{\openn}{\hbox{${\rm I\kern-.2em N}$}}

\endlocaldefs

\usepackage{natbib}
\usepackage{colortbl}
\usepackage{lipsum}
\usepackage{amsmath,amsfonts,amsthm,amssymb,mathtools}
\usepackage{mathrsfs}
\usepackage{indentfirst}

\usepackage{ifthen}
\usepackage{xifthen}
\DeclarePairedDelimiterX{\bracks}[1]{\lbrack}{\rbrack}{#1}
\DeclarePairedDelimiterX{\set}[1]{\lbrace}{\rbrace}{#1}
\newcommand{\ex}{\mathbb{E}}
\newcommand{\expect}[2][] {
    \ifthenelse{\isempty{#1}}{
        \ex\bracks*{#2}
    }{
      \ex_{#1}\bracks*{#2}  
    }
}
\newcommand{\indicator}[1]{\mathbb{I}_{\set*{#1}}}
\usepackage{xcolor}
\usepackage{soul}
\usepackage{arydshln}
\definecolor{LightCyan}{rgb}{0.88,1,1}
\newcommand{\newbrace}[1]{\left\{#1\right\}}
\newcommand{\newbracket}[1]{\left[#1\right]}
\newcommand{\newparethensis}[1]{\left(#1\right)}
\newcommand\numberthis{\addtocounter{equation}{1}\tag{\theequation}}
\newcommand{\covn}[2][] {
    \ifthenelse{\isempty{#1}}{
        {\textrm{cov}_n}\parens*{#2}
    }{
      {\textrm{cov}}_{#1}\parens*{#2}  
    }
}
\newcommand{\varn}[2][] {
    \ifthenelse{\isempty{#1}}{
        {\textrm{var}_n}\parens*{#2}
    }{
      {\textrm{var}}_{#1}\parens*{#2}  
    }
}

\newcommand{\norm}[1]{\left\lVert#1\right\rVert}

\usepackage{comment}
\usepackage{algorithm} 
\usepackage{algpseudocode}

\newcommand{\parents}{
\text{Pa}
}

\newcommand{\children}{
\text{Ch}
}

\begin{document}

\begin{frontmatter}
\title{Targeted Maximum Likelihood Based Estimation for Longitudinal Mediation Analysis}
\runtitle{TMLE for Longitudinal Mediation}

\begin{aug}
\author[A]{\fnms{Zeyi} \snm{Wang}\ead[label=e1,mark]{wangzeyi@berkeley.edu}}, 
\author[B]{\fnms{Lars} \snm{van der Laan}\ead[label=e2]{vanderlaanlars@yahoo.com}}, 
\author[A]{\fnms{Maya} \snm{Petersen}\ead[label=e3,mark]{mayaliv@berkeley.edu}},
\author[C]{\fnms{Thomas} \snm{Gerds}\ead[label=e5]{tag@biostat.ku.dk}}, 
\author[D]{\fnms{Kajsa} \snm{Kvist}\ead[label=e6]{tekk@novonordisk.com}},
\and
\author[A]{\fnms{Mark} \snm{van der Laan}\ead[label=e4,mark]{laan@stat.berkeley.edu}},

\address[A]{Division of Biostatistics, School of Public Health, University of California, Berkeley, Berkeley, USA, 
\printead{e1,e3,e4}}

\address[B]{Division of Environmental Health Sciences, School of Public Health, University of California, Berkeley, Berkeley, USA, 
\printead{e2}}

\address[C]{Section of Biostatistics, Department of Public Health, University of Copenhagen, Copenhagen, Denmark, 
\printead{e5}}

\address[D]{Novo Nordisk, Søborg, Denmark,
\printead{e6}}
\end{aug}

\begin{abstract}
Causal mediation analysis with random interventions has become an area of significant interest for understanding time-varying effects with longitudinal and survival outcomes. 
To tackle causal and statistical challenges due to the complex longitudinal data structure with time-varying confounders, competing risks, and informative censoring, there exists a general desire to combine machine learning techniques and semiparametric theory. 
In this manuscript, we focus on targeted maximum likelihood estimation (TMLE) of longitudinal natural direct and indirect effects defined with random interventions. 
The proposed estimators are multiply robust, locally efficient, and directly estimate and update the conditional densities that factorize data likelihoods.
We utilize the highly adaptive lasso (HAL) and projection representations to derive new estimators (HAL-EIC)  of the efficient influence curves of longitudinal mediation problems and propose a fast one-step TMLE algorithm using HAL-EIC while preserving the asymptotic properties. 
The proposed method can be generalized for other longitudinal causal parameters that are smooth functions of data likelihoods, 
and thereby provides a novel and flexible statistical toolbox. 
\end{abstract}



\end{frontmatter}

\section{Introduction}

There is an increasing need of methods that can analyze mechanisms of temporally varying effects, for example, in clinical trials and studies of electronic health record data \citep{vansteelandt2019mediation,buse2020cardiovascular,lai2020understanding}.  For example, a weight management program may involve repeated scheduled visits where measurements of body mass index (BMI), blood pressure, cholesterol levels, and other health conditions are collected. One may want to analyze the proportion of the effect of weight loss on cholesterol (continuous outcome) or risks of cardiovascular events (time-to-event outcomes) that is mediated by how well blood pressure is under control. It may further be of interest to study how this proportion changes over time. Mediation analysis provides estimands that address these questions. However, state of the art static intervention based mediation analysis is limited in the ability of properly adjusting for time-varying confounding without additional model assumptions, see the discussion of natural effects being nonparametrically non-identifiable in presence of ``recanting witnesses'' in \cite{avin2005identifiability}.
Generally, longitudinal mediation is a challenging problem \citep{avin2005identifiability,vanderweele2009conceptual,tchetgen2012semiparametric,zheng2012causal,zheng2017longitudinal,vanderweele2017mediation} and the task to adjust for  time-varying confounding
is further complicated when the longitudinal causal mediation target parameters, such as natural direct and indirect effects, are defined with static interventions \citep{vanderweele2017mediation}.

Recently, mediation analysis based on random intervention has been proposed \citep{didelez2006direct,diaz2020causal,vanderweele2017mediation,zheng2012causal,zheng2017longitudinal}. Instead of deterministically enforcing specific mediator values in structural causal models \citep{pearl2009causal}, the random interventions define causal targets by enforcing mediator distributions. This allows flexible time-varying confounding adjustment, whereas nonparametric identification fails for natural effects using static interventions with the existence any post-treatment mediator-outcome confounders \citep{avin2005identifiability}. 
In the non-longitudinal settings without mediator-outcome confounders impacted by intervention (the so-called recanting witnesses), random intervention targets can be identified with the same g-computation formulas known from the corresponding classical static interventions. However, random intervention targets allow us to relax the cross-world identification assumptions for natural direct and indirect effects (NDE and NIE), i.e., the untestable conditional independence between counterfactual mediators and outcomes defined with different static interventions \citep{robins2010alternative}.

In this manuscript, we develop the targeted likelihood based method \citep{van2010targeted,van2010targeted2} for longitudinal mediation parameters and construct targeted maximum likelihood estimators (TMLEs). We derive conditions under which the TMLEs become consistent and asymptotically linear. We also provide a projection representation (HAL-EIC) for the efficient influence curves for longitudinal mediation problems and use it to derive a fast one-step TMLE algorithm. Throughout the manuscript we will focus on the longitudinal analysis of NIE and NDE, but the methodology immediately applies to controlled direct effects (CDE) (see Appendix \ref{sec:CDE_EIC}). This flexibility is due to that the g-computation formulas for these target parameters can be represented as functions of the same observed data likelihood. To explain the main idea of our targeted likelihood based method for multivariate target parameters, consider a discrete time scale with \(K\) time points. Let $O = (X_0, X_1, \dots, X_K)$ denote the observed data and let $O_1, \dots, O_n \sim P_0$ be an IID sample, and let $p = dP/d\mu$ be the density function with respect to a dominating measure for a distribution $P$. A g-computation formula can be labeled by the counterfactual distribution $P(g)$ of $O$ under an intervention $g$, where $P(g)$ is identified as a function of the factorized observed data likelihood $p_0(O) = p_{0, X_0}(X_0) \prod_{k = 1}^K p_{0, X_k}(X_k | X_0, \dots, X_{k-1})$. 
Suppose that the aim of the analysis is not just a single target parameter but a set of target parameters $\{\Psi_s(P_0), s = 1, \dots, S\}$ where \(S\) is the total number of target parameters. This occurs when there are multiple interventions and/or multiple endpoints are of interest such that each combination of  an intervention and an outcome defines a target parameter. We require each target parameter to be pathwise differentiable and denote $D_s(P)$ for the efficient influence curve of \(\Psi_s\). The list of target parameters could for example contain NIE and NDE for more than one intervention and for a sequence of evaluation time points. Our two-stage targeted likelihood based estimation approach thus starts with an initial estimate of the full likelihood $p^0_n$ of $p_0$, and then searches for an updated estimate of the likelihood $p_n^*$ which solves the efficient influence curve equations $P_n D_s(p_n^*)=0, s = 1, \dots, S$ of all target parameters simultaneously. We show in this manuscript that the plug-in estimators $\Psi_s(p_n^*), s = 1, \dots, S$ (TMLEs) derived from the same updated estimate of the likelihood $p_n^*$ are consistent and asymptotically efficient. We also show that the TMLEs respect the parameter space of all targets simultaneously, that is, in the sense of \cite{robins2007comment} the estimates stay in the range of all parameter mappings implied by the statistical model .

We argue that our unified likelihood-based approach for longitudinal mediation parameters has advantages compared to recent work \citep{vanderweele2017mediation} that focuses on restricted marginal models where natural effects no longer decompose the total effect and are not readily suitable for survival outcomes, or sequential regression components \citep{zheng2017longitudinal}, with iteratively defined conditional expectations similar to \cite{bang2005doubly}, that are not shared across different target parameters.
As pointed out above, our full likelihood-based approach is natural and flexible for simultaneous analysis across multiple time points and multiple targets. In settings with time-varying confounders, survival outcomes, informative right-censoring, and competing risks, the plug-in estimators derived from the same updated likelihood will respect the parameter space of all the targets. 
The sequential regression may significantly reduce the computational cost of estimation and may also respect the parameter space \citep{robins2007comment} but only in each of its dimensions marginally. 
 This makes it more difficult to conduct loss-based collaborative adjustment with a sequence of candidate propensity score models and the corresponding TMLE updates
\citep{van2010collaborative}. With likelihood based estimation, the loss and adjustment procedure can be naturally defined even for multivariate targets by minimizing negative log-likelihood losses of the same set of conditional density factors. 
Identification assumptions and multiple robustness conditions are more communicable when the targets of inference are identified and represented as functions of the observed data distributions without additional iterated definitions  and unintuitive restrictions. We note that the conditions for multiple robustness in Section \ref{sec:MR} involve only conditional densities of observed data, and thus our approach avoids the need to specify models for sequential regression. Simple and direct robustness conditions are more suitable for conceptual verification, which lead to more effective use of experts' data knowledge and improved reliability of research.

This manuscript presents a novel contribution to the field of longitudinal mediation analysis through the proposed HAL-EIC representation and the development of a fast one-step TMLE algorithm using HAL-EIC. 
The highly adaptive lasso (HAL) \citep{benkeser2016highly,bibaut2019fast,van2018highly} is a general maximum likelihood estimator (MLE) for $d$-variate real-valued cadlag functions bounded in the sectional variation norm, which converges to the true function at a rate of $n^{-2/3}(\log n)^d$ in loss-based dissimilarity. Recent developments \citep{van2018hal} have demonstrated that HAL estimators offer a promising alternative representation of efficient influence curves (EIC) for many pathwise differentiable target parameters. In this manuscript, we propose the use of the HAL-EIC representation for longitudinal mediation problems and develop a fast  one-step TMLE algorithm using HAL-EIC while preserving the asymptotic properties.

The outline of the manuscript is as follows. Section \ref{sec:data} introduces the notation and the data structure for discrete time mediation analysis with outcomes that may be suitable for the proposed method, such as survival outcomes, right censoring, and competing risks. Section \ref{sec:target} defines the causal mediation target parameters, and identifies them as statistical targets using g-computation formulas, and provides the efficient influence curves. Section \ref{sec:initial}-\ref{sec:tmle} discusses the two-stage TMLE procedure. In Section \ref{sec:initial} we briefly discuss the steps and considerations for constructing initial density estimators. In Section \ref{sec:tmle} we give the iterative and one-step TMLE algorithms \citep{van2016one} and discuss the regularity conditions for local efficiency. We review a method of simultaneous inference on multivariate target parameters introduced by \cite{dudoit2008multiple} in our setting. In Section \ref{sec:num}, we propose a projection representation (HAL-EIC) for the efficient influence curves using the highly adaptive lasso (HAL) \citep{benkeser2016highly,hejazi2020hal9001}, and propose the HAL-EIC based one-step TMLE algorithm. In the following sections, we analyze the multiple robustness properties \citep{tchetgen2009commentary,molina2017multiple,luedtke2017sequential,zheng2017longitudinal} using simulated data and show results of the proposed algorithms under finite-sample challenges such as near-violation of the positivity assumptions. 

\section{Data and Model}\label{sec:data}

On a discrete time scale, \(\{0,1,\dots,K\}\), we denote for all \(0\le t\le K\) by
$\bar X_t$ a vector of random variables measured until time \(t\), i.e., $\bar X_t=(X_k: k \in \{0, \dots, t\})$.  Similarly, for $k \leq t$ we define $\bar X_{k}^t = (X_k, \dots, X_t)$. 
Throughout,  we consider the following data structure:
\begin{equation*}
    O  = \newparethensis{L_0, A_1, R_1, Z_1, L_1, \ldots, A_t, R_t, Z_t, L_t\ldots, A_K, R_K, Z_K, L_K}, 
\end{equation*}
where each node is a random variable or a random vector. $L_0$ are the baseline covariates, $\bar A_K$ are treatment variables, $\bar Z_K$ are time-varying mediators, $\bar R_K$ and $\bar L_K$ are time-varying variables before and after the mediator nodes. The vector $(A_t, R_t, Z_t, L_t)$ is the observation at time point $t$. Given this time order, we denote by $\parents(X_t)$ the parent variables preceding $X_t$, and by $\children(X_t)$ the child variables after $X_t$. Similar notations are used for parent and child nodes of a realization $o = (l_0, a_1 \dots, l_K)$ in the range of $O$. For example, $\parents(L_1) = (L_0, A_1, R_1, Z_1)$, whereas $\children(a_K) = (r_K, z_K, l_K)$. In the context of a static intervention $\bar a_K$ which sets treatment values, we denote by \(\parents(X_t|\bar{a}_K)\) the parent nodes of \(X_t\) under the intervention \(\bar{A}_K=\bar{a}_K\), e.g., 
$\parents(L_k | \bar a_k) = (L_0, a_1, R_1, Z_1, L_1, \dots, a_k, R_k, Z_k)$, which is an ordered vector of observed random variables and imputed treatment values. 
The outcome of interest is defined as a vector of functions of $\bar L_K$, that is, $\bar Y = \bar \psi(\bar L_K)$ for a multi-dimensional functional $\bar \psi = (\psi_s: s = 1, \dots, S)$. Examples include cumulative measures or survival status at different time points. A simple special case is a univariate variable \(Y\) measured at the last time point such that $Y \in L_K$.  

Let \(\mathcal{M}\) be a statistical model for the distribution of \(O\) which is dominated by a measure \(\mu\) such that for \(P\in\mathcal{M}\) the density is given by $p = dP/d\mu$. 
For clarity of representation we assume $\mu$ to be a counting measure so that integrals such as (\ref{eq:gcomp}) and (\ref{eq:rewrite}) are simplified as summations, but the results generalize to continuous probability distributions dominated by Lebesgue measures and corresponding integrals. 
We identify $P \in \mathcal{M}$ with the corresponding density \(p\), and treat distributions with identical densities as equivalent.
We consider $n$ IID copies $O_1, \ldots, O_n$ of $O$ that follow the true data generating distribution $P_0\in\mathcal{M}$. The joint data likelihood can be factorized by the conditional densities, 
\begin{multline}
p(O) = p_{L_0}(L_0) \prod_{k = 1}^K \Big[
 p_{A_k}(A_k | \parents(A_k)) 
 p_{R_k}(R_k | \parents(R_k)) \\
 p_{Z_k}(Z_k | \parents(Z_k)) 
 p_{L_k}(L_k | \parents(L_k)) \Big].
\label{eq:full_lkd}
\end{multline}
We use the notation $Pf(O) = \ex_P\newbracket{f(O)}$ for the expectation with respect to \(P\) and likewise $P_n f(O) = \frac{1}{n} \sum_{i=1}^n f(O_i)$ for the empirical distribution $P_n$. 

\subsection{Structural Causal Model and Random Interventions} \label{sec:RI}

Consider the following structural causal model \citep{pearl2009causal,van2011targeted}, where the randomness of the observed data $O$ is captured by the so-called exogenous nodes $U_X$'s, and the $f_{X}$ mappings are deterministic for each of the endogenous variables:
\begin{equation*}
    X_t=f_{X_t}(\parents(X_t),U_{X_t})
\end{equation*}
For some static treatment $\bar A_K = \bar a_K$, we define counterfactuals by static interventions on the structural causal model. 
\begin{align*}
    L_0 = & f_{L_0}(U_{L_0}), \quad A_t = a_t \\
    R_t(\bar a) = & f_{R_t}(\bar a_{t}, \bar R_{t-1}(\bar a), \bar Z_{t-1}(\bar a), \bar L_{t-1}(\bar a), U_{R_t}) \\
    Z_t(\bar a) = & f_{Z_t}(\bar a_{t}, \bar R_{t}(\bar a), \bar Z_{t-1}(\bar a), \bar L_{t-1}(\bar a), U_{Z_t}) \\
    L_t(\bar a) = & f_{L_t}(\bar a_{t}, \bar R_{t}(\bar a), \bar Z_{t}(\bar a), \bar L_{t-1}(\bar a), U_{L_t}).
\end{align*}

For a pair of static treatment and control interventions, $\bar a$ and $\bar a'$, denote by $\Gamma_t^{\bar a'}$ the conditional density of the control intervention counterfactual $Z_t(\bar a')$ given parent nodes under $\bar a'$, that is, $\Gamma_t^{\bar a'}(z_t | \bar r_t, \bar z_{t-1}, \bar l_{t-1}) = p(Z_t(\bar a') = z_t | \bar R_{t}(\bar a') = \bar r_t, \bar Z_{t-1}(\bar a') = \bar z_{t-1}, \bar L_{t-1}(\bar a') = \bar l_{t-1})$. The vector $\bar \Gamma^{\bar a'} = \newparethensis{\Gamma_t^{\bar a'}: t = 1, \dots, K}$ thus represents a sequence of  counterfactual conditional densities for the mediator process under the control intervention.
We define random intervention counterfactuals under $(\bar a, \bar \Gamma^{\bar a'})$ by forcing mediator variables to follow the distributions of the control intervention counterfactuals, $Z_t(\bar a, \bar \Gamma^{\bar a'}) \sim \Gamma_t^{\bar a'}$, and for nodes $X_t \in \bar R \cup \bar L$ inserting $\bar A_t = \bar a_t$ so that $X_t(\bar a, \bar \Gamma^{\bar a'}) = f_{X_t}(\parents(X_t(\bar a, \bar \Gamma^{\bar a'}) | \bar a), U_X)$. That is, 
\begin{align*}
    L_0(\bar a, \bar \Gamma^{\bar a'}) = & f_{L_0}(U_{L_0}), \quad A_t = a_t \\
    R_t(\bar a, \bar \Gamma^{\bar a'}) = & f_{R_t}(\bar a_{t}, \bar R_{t-1}(\bar a, \bar \Gamma^{\bar a'}), \bar Z_{t-1}(\bar a, \bar \Gamma^{\bar a'}), \bar L_{t-1}(\bar a, \bar \Gamma^{\bar a'}), U_{R_t}) \\
    Z_t(\bar a, \bar \Gamma^{\bar a'}) \sim &  \Gamma_t^{\bar a'}(z | \bar R_{t}(\bar a, \bar \Gamma^{\bar a'}), \bar Z_{t-1}(\bar a, \bar \Gamma^{\bar a'}), \bar L_{t-1}(\bar a, \bar \Gamma^{\bar a'}))\\
    L_t(\bar a, \bar \Gamma^{\bar a'}) = & f_{L_t}(\bar a_{t}, \bar R_{t}(\bar a, \bar \Gamma^{\bar a'}), \bar Z_{t}(\bar a, \bar \Gamma^{\bar a'}), \bar L_{t-1}(\bar a, \bar \Gamma^{\bar a'}), U_{L_t}). 
\end{align*}

In this framework, a causal target parameter that describes the mediation of treatment effects on an outcome $Y$ is denoted by $\expect{Y(\bar a, \bar \Gamma^{\bar a'})}$. Then the decomposition of the total effect into a natural indirect effect (NIE) and a natural direct effect (NDE) is given by 
\begin{multline*}
    \expect{Y(\bar a, \bar \Gamma^{\bar a})} - \expect{Y(\bar a', \bar \Gamma^{\bar a'})}
    = 
    \newparethensis{\expect{Y(\bar a, \bar \Gamma^{\bar a})} - \expect{Y(\bar a, \bar \Gamma^{\bar a'})}} \\ + 
    \newparethensis{\expect{Y(\bar a, \bar \Gamma^{\bar a'})} - \expect{Y(\bar a', \bar \Gamma^{\bar a'})}}. 
\end{multline*}

In order to explain the cross-world assumptions, we also define counterfactuals $X(\bar a, \bar z)$ with static interventions $\bar A = \bar a$ and $\bar Z = \bar z$. Note that if after a random draw under $(\bar a, \bar \Gamma^{\bar a'})$ we have that $\bar Z(\bar a, \bar \Gamma^{\bar a'}) = \bar z$, then the structural causal model implies $X(\bar a, \bar \Gamma^{\bar a'}) = X(\bar a, \bar z)$.
However, defining NIE and NDE without random intervention usually requires additional assumptions such as conditional independence between $L(\bar a, \bar z)$ and $Z(\bar a')$ across the two counterfactual worlds. Such ``cross-world'' assumptions are not desirable because they are not verifiable, neither empirically nor conceptually, see \cite{andrews2020insights}. Cross-world assumptions are also incompatible with time-varying covariates $\bar R$. On the other hand, the random intervention framework does not require ``cross-world'' assumptions. 
In this manuscript, we focus on static treatment rules $\bar A = \bar a$ or $\bar a'$, but the methodology can be generalized to dynamic treatment regimes, where the treatment and control interventions on $A_t$ are decided by deterministic functions of the available history $\parents(A_t)$.

\subsection{Right Censored Survival Outcomes}\label{sec:survival}

Suppose $T$ is the time point where an event of interest happens. Let $Y_t = \indicator{T > t} \in L_t$ be the monotone process of staying event-free in the study at the $t$-th time point. 
The counting process $Y_t$ starts with $1$ at $t=0$ and only jumps to $0$ if an event happens at the time point $t$. If an event happens at the time point $t$, then for all $X \in \children(Y_t) \setminus \bar Y$, conditional on $Y_t = 0$ we set $X$ to be a degenerated discrete variable such that  $X = \varnothing$ with conditional probability $1$. The outcome of interest can be the survival beyond the study length $K$, in which case the target parameters take the form of 
\begin{align*}
\expect{Y_K(\bar a, \bar{\Gamma}^{\bar a'})} = P(T(\bar a, \bar{\Gamma}^{\bar a'}) > K). 
\end{align*}

In real applications, one typically only observes $\Tilde{T} = \min\newbrace{T, C}$, where $C$ is the censoring time. To incorporate censored data, we create bivariate treatment nodes $A_t = (A_t^C, A_t^E)$ that consist of  the monotone process of remaining uncensored, $A_t^C = \indicator{C > t}$, and the treatment assignment, \(A_t^E\). The process $A_t^C$ starts with value $1$ and jumps to $0$ at the time point where censoring occurs.
Conditioning on $C_t = 0$, for all $X \in \children(C_t)$, we set $X$ to be a discrete variable that equals a degenerated value $\varnothing$ with conditional probability $1$; i.e., no information is available for the observed data likelihood after censoring occurs. 

\subsection{Competing Risks}\label{sec:competing}

Consider a competing events framework \citep{abgk,benkeser2018improved,rytgaard2022targeted} where at the time \(T\) where an individual reaches one of several absorbing states the event type \(\Delta\in\{1,2,\dots,J\}\) is observed.  For example, $\Delta=1$ may indicate the onset of a cancer, and $\Delta=2$ death due to other causes. One can define a multi-dimensional counting process in discrete time, e.g., when \(J=2\) by $\newbrace{Y_t = (N_t^{(1)}, N_t^{(2)}): t = 1, \dots, K}$, such that $ N^{(1)}_t = \indicator{T \leq t, \Delta=1)}$, $N^{(2)}_t = \indicator{T \leq t, \Delta=2)}$. Note that \(1 - N_t^{(1)} - N_t^{(2)}=\indicator{T>t}\) now indicates that the individual is alive and event-free. Suppose $Y_t \in L_t$. For $Y_t = (0, 0)$, we set $X = \varnothing$ for any $X \in \children(Y_t)\setminus\bar Y$ with probability $1$. If $N^{(j)}_t = 1$ for the $j$-th type of event, we additionally set $N^{(j)}_{t'} = 1$ and $N^{(j')}_{t'} = \varnothing$ for all $j'\not=j, t'\geq t$. 
The target parameter in a competing risk framework is typically multi-dimensional and can for example be the risks of all events under $(\bar a, \bar \Gamma^{\bar a'})$ across all time points, $\ex\newbracket{\bar Y(\bar a, \bar \Gamma^{\bar a'})} =  \newparethensis{\ex\newbracket{N^{(j)}_t(\bar a, \bar \Gamma^{\bar a'})}: t = 1, \dots, K; j=1,\dots, J}$. 

\section{Target Parameters}\label{sec:target}

\subsection{Natural Direct and Indirect Effects}

We define natural direct effects (NDE) and natural indirect effects (NIE) for a  multi-dimensional outcome of interest $\bar Y = \bar \psi(\bar L_K)$ with random interventions as defined in Section \ref{sec:data}: 
\begin{align*}
\text{NIE: } & \expect{\bar Y(\bar a, \bar \Gamma^{\bar a}) - \bar Y(\bar a, \bar \Gamma^{\bar a'})},\\
\text{NDE: } & \expect{\bar Y(\bar a, \bar \Gamma^{\bar a'}) - \bar Y(\bar a', \bar \Gamma^{\bar a'})}.
\end{align*}
Examples for the choice of $\bar Y$ include: $Y \in L_K$ for an univariate outcome variable, see Section $\ref{sec:RI}$, $Y_K = \indicator{T > K} \in L_K$ for censored survival endpoints, see Section $\ref{sec:survival}$, and $\bar{Y} = \newparethensis{N^{(1)}_t, N^{(2)}_t: t = 1, \dots, K}$ for competing risk outcomes, see Section \ref{sec:competing}. 

NDE is the direct effect of a treatment while forcing mediators to have the same distribution as their control group counterfactuals. NIE is the indirect treatment effect achieved by not changing treatment values but by changing mediator distributions. 
The structural causal model of Section \ref{sec:RI} implies that NDE and NIE decompose the total effect, which can be defined with or without random intervention:
$$\text{NIE} + \text{NDE} = \expect{\bar Y(\bar a, \bar \Gamma^{\bar a})} - \expect{\bar Y(\bar a', \bar \Gamma^{\bar a'})} = \expect{\bar Y(\bar a)} - \expect{\bar Y(\bar a')}.$$

\subsection{Identification} \label{sec:id}

To identify the  causal mediation targets as statistical parameters of the observed data distribution, we adopt the following assumptions from \cite{zheng2017longitudinal}. For any random intervention $(\bar a, \bar \Gamma^{\bar a'})$ of interest: 
\begin{enumerate}[label=(A\arabic*)]
    \item\label{cond:seq_ex} Sequential exchangeability: $\bar R_{t}^{K}(\bar a'), \bar Z_{t}^{K}(\bar a'), \bar L_{t}^{K}(\bar a'), \bar R_{t}^{K}(\bar a, \bar z), \bar L_{t}^{K}(\bar a, \bar z) \perp A_t | \parents(A_t)$. 
    \item Mediator randomization: $\bar R_{t+1}^{K}(\bar a, \bar z), \bar L_{t}^{K}(\bar a, \bar z) \perp Z_t | \parents(Z_t)$. 
    \item\label{cond:pos} (Strong) Positivity: 
    $p_0(a^*_t | \parents(a^*_t | \bar a^*)) > 0$ if $p_0(\parents(a^*_t | \bar a^*)) > 0$ for $\bar a^* = \bar a$ or $\bar a'$; also, 
    $p_0(r_t | \parents(r_t | \bar a'_t)) > 0$ if $p_0(r_t | \parents(r_t| \bar a_t)) > 0$, 
    $p_0(l_t | \parents(l_t| \bar a'_t)) > 0$ if $p_0(l_t | \parents(l_t| \bar a_t)) > 0$, and
    $p_0(z_t | \parents(z_t| \bar a_t)) > 0$ if $p_0(z_t | \parents(z_t| \bar a'_t)) > 0$. 
\end{enumerate}
Note that the consistency assumption \citep{cole2009consistency} is implicitly made via the structural causal model. 
The following theorem is a direct application of Lemma 1 in \cite{zheng2017longitudinal} to multivariate outcome $\bar Y = \newparethensis{Y_s: s}$ on each dimension. 
\begin{theorem}[G-computation formula]
Suppose that the multi-dimensional outcome of interest is $\bar Y = (Y_s: s = 1, \dots, S)$ where $Y_s = \psi_s(\bar L_K)$ is defined by measurable functions $\psi_s$.
For any two interventions $\bar a, \bar a'$ under the identification assumptions \ref{cond:seq_ex}-\ref{cond:pos} listed in Section \ref{sec:id} we have
\begin{align*}
    \Psi_s^{\bar a, \bar a'}(P)
    \equiv & \expect{Y_s(\bar a, \bar{\Gamma}^{\bar a'})} \\
    = & \sum_{\bar l} \psi_s(\bar l_K) p(\bar L_K(\bar a, \bar \Gamma^{\bar a'}) = \bar l_K) \\
    = & \sum_{\bar l, \bar z, \bar r} \psi_s(\bar l_K) p(L_0 = l_0)  
     \prod_{t = 1}^K p_{L_t}(l_t | \parents(l_t| \bar a_t))
     p_{Z_t}(z_t | \parents(z_t| \bar a'_t))
     p_{R_t}(r_t | \parents(r_t| \bar a_t)). \numberthis \label{eq:gcomp}
\end{align*}
Also, for NIE and NDE w.r.t. each of the dimensions of the outcome, 
\begin{align*}
\text{NIE}_s(P)
= & \Psi_s^{\bar a, \bar a}(P) - \Psi_s^{\bar a, \bar a'}(P) \\
\text{NDE}_s(P)
= & \Psi_s^{\bar a, \bar a'}(P) - \Psi_s^{\bar a', \bar a'}(P).
\end{align*}
\end{theorem}

\subsection{Comparison with Sequential Regression}\label{sec:sr}

We have defined and identified our target parameters as functions of the observed data likelihood $p$ and we will focus on such full likelihood representation in the following sections. In this subsection, we compare with the sequential regression (or iterated conditional expectation) approach \citep{bang2005doubly,zheng2017longitudinal} which rewrites the same g-computation formula as functions of iteratively defined regression components. 

For any pair of interventions $\bar a$ and $\bar a'$, one can identify the post-intervention distribution $P^{\bar a, \bar a'}$ of the counterfactual nodes under random interventions $\newparethensis{\bar a, \bar \Gamma^{\bar a'}}$ given the assumptions in Section \ref{sec:id}. That is, for any data realization $o$ that enforces a value $\bar a_K$ in the intervention nodes, we have
\begin{align*}
    p^{\bar a, \bar a'}(O = o) 
    = & p(O(\bar a, \Gamma^{\bar a'}) = o) \\
    = & p(\bar L_K(\bar a, \Gamma^{\bar a'}) = \bar l_K, \bar Z_K(\bar a, \Gamma^{\bar a'}) = \bar z_K, \bar R_K(\bar a, \Gamma^{\bar a'}) = \bar r_K) \\
    = & \prod_{t = 1}^K p_{L_t}(l_t | \parents(l_t| \bar a_t))
     p_{Z_t}(z_t | \parents(z_t| \bar a'_t))
     p_{R_t}(r_t | \parents(r_t| \bar a_t)). 
\end{align*}
For each dimension \(s\) of the statistical target parameter, one can rewrite the g-computation formulas with the following sequence of regression expressions (the dependence of $Q$ on $P$ is suppressed except for the last equation): 
\begin{align*}
    Q_{s, R_{K + 1}}^{\bar a, \bar a'} = & Y_s = \psi_s(\bar L_K) \\
    Q_{s, L_t}^{\bar a, \bar a'}\newparethensis{\bar R_t, \bar Z_t, \bar L_{t-1}} = & \ex_{P^{\bar a, \bar a'}}\newbracket{Q_{s, R_{t+1}}^{\bar a, \bar a'} | \bar R_t, \bar Z_t, \bar L_{t-1}} = \ex_{P}\newbracket{Q_{s, R_{t+1}}^{\bar a, \bar a'} | \bar R_t, \bar Z_t, \bar L_{t-1}, \bar A_t = \bar a_t} \\
    Q_{s, Z_t}^{\bar a, \bar a'}\newparethensis{\bar R_t, \bar Z_{t-1}, \bar L_{t-1}} = & \ex_{P^{\bar a, \bar a'}}\newbracket{Q_{s, L_{t}}^{\bar a, \bar a'} | \bar R_t, \bar Z_{t-1}, \bar L_{t-1}} = \ex_{P}\newbracket{Q_{s, L_{t}}^{\bar a, \bar a'} | \bar R_t, \bar Z_{t-1}, \bar L_{t-1}, \bar A_t = \bar a'_t} \\
    Q_{s, R_t}^{\bar a, \bar a'}\newparethensis{\bar R_{t-1}, \bar Z_{t-1}, \bar L_{t-1}} = & \ex_{P^{\bar a, \bar a'}}\newbracket{Q_{s, Z_{t}}^{\bar a, \bar a'} | \bar R_{t-1}, \bar Z_{t-1}, \bar L_{t-1}} \\ &= \ex_{P}\newbracket{Q_{s, Z_{t}}^{\bar a, \bar a'} | \bar R_{t-1}, \bar Z_{t-1}, \bar L_{t-1}, \bar A_t = \bar a_t} \\
    \Psi_s^{\bar a, \bar a'}(P) = & \ex_{P} Q_{s, R_1}^{\bar a, \bar a'}(P)(L_0). 
\end{align*}
Note that each of the regression expressions can also be written as a function of conditional densities by analytically carrying out the nested conditional expectations. For example, for simplicity assuming that $L_k$ is a discrete variable we have
\begin{multline}
    Q_{s, L_K}^{\bar a, \bar a'} = \ex_{P}\newbracket{Q_{s, R_{K+1}}^{\bar a, \bar a'} | \bar R_K, \bar Z_K, \bar L_{K-1}, \bar A_t = \bar a_t}\\
= \sum_{l_K} \psi_s(\bar L_{K-1}, l_K) p_{L_K}(l_k | \bar R_K, \bar Z_K, \bar L_{K-1}, \bar A_K = \bar a_K). \label{eq:rewrite}
\end{multline}

The sequential regression formulation thus provides an alternative representation of the g-computation formula of $\Psi_s^{\bar a, \bar a'}(P)$ as a functional of 
\begin{equation*}\bar Q^{\bar a, \bar a'}_s(P) = \newparethensis{Q_{s, X}^{\bar a, \bar a'}(P): X \in \newbrace{L_t, Z_t, R_t: t = 1, \dots, K}}.\end{equation*}
It requires only estimating regression models, $Q_{s, X}^{\bar a, \bar a'}(P)$, of $P$ instead of the factorized data likelihood (\ref{eq:full_lkd}), which leads to a potential gain in scalability. However, this comes at a cost of capturing less information from the data distribution $P$ and only works for one target parameter at a time. Having to define and estimate a different set of intermediate regression components $\bar Q_{s}^{\bar a, \bar a'}(P)$ for each dimension $s$ of the outcome of interest and each random intervention $(\bar a, \Gamma^{\bar a'})$ of interest is not intuitive and it potentially leads to results which do not obey the parameter space. While the resulting estimates still respect the marginal parameter spaces for each dimension of the target parameter due to substitution estimation, sequential regression cannot guarantee the ``boundedness'' property \citep{robins2007comment} for the joint parameter space. The target parameter is usually multi-dimensional in longitudinal mediation analysis. For example, when the outcome of interest is the whole survival process $\bar Y = (Y_t = \indicator{T > t}: t = 1, \dots, K)$, the estimations of $\expect{Y_t(\bar a, \bar \Gamma^{\bar a'})}$ are expected to be monotonic in $t = 1, 2, \dots, K$. Another example is the competing risks in Section \ref{sec:competing}, where the estimations also need to respect that $\expect{Y_t^{(1)}(\bar a, \bar \Gamma^{\bar a'})} + \expect{Y_t^{(2)}(\bar a, \bar \Gamma^{\bar a'})} = \expect{Y_t^*(\bar a, \bar \Gamma^{\bar a'})} \in \newbracket{0, 1}$, increasing in $t = 1, \dots, K$. 
Our TMLE algorithms on the other hand derive estimates of all target parameters from the same set of estimated likelihood factors and hence achieve the desired joint boundedness property. 

\subsection{Efficient Influence Curve (EIC)}\label{sec:EIC}

\begin{theorem}[Efficient influence curve (EIC)]
For each pair of interventions $\bar a, \bar a'$ and the $s$-th element of the outcome vector $\bar Y$, the efficient influence curve $D^{\bar a, \bar a'}_{s}(P)$ of $\Psi^{\bar a, \bar a'}_s$ at $P$ is given by (some dependence on $P$ is suppressed): 
\begin{align*}
D^{\bar a, \bar a'}_{s, L_t} = & \frac{\indicator{\bar A_t = \bar a_t}}{\prod_{j = 1}^t p_{A}(a_j | \parents(A_j| \bar a_{j-1}))}
   \\&\qquad\qquad\qquad \prod_{j = 1}^t \frac{p_Z(Z_j | \parents(Z_j| \bar a'_j))}{p_Z(Z_j | \parents(Z_j| \bar a_j))} 
    \newbrace{Q_{s, R_{t+1}}^{\bar a, \bar a'}(\bar R_t, \bar Z_t, \bar L_{t}) - 
    Q_{s, L_{t}}^{\bar a, \bar a'}(\bar R_t, \bar Z_t, \bar L_{t-1})} \\
D^{\bar a, \bar a'}_{s, Z_t} = & \frac{\indicator{\bar A_t = \bar a_t'}}{\prod_{j = 1}^t p_A(a_j' | \parents(A_J| \bar a'_{j-1}))}
    \prod_{j = 1}^{t-1} \frac{p_L(L_j | \parents(L_j| \bar a_j))}{p_L(L_j | \parents(L_j| \bar a_j'))}
    \prod_{j = 1}^t \frac{p_R(R_j | \parents(R_j| \bar a_j))}{p_R(R_j | \parents(R_j| \bar a_j'))} \\
    & \newbrace{Q_{L_{t}}^{\bar a, \bar a'}(\bar R_t, \bar Z_{t}, \bar L_{t-1}) - 
    Q_{Z_{t}}^{\bar a, \bar a'}(\bar R_t, \bar Z_{t-1}, \bar L_{t-1})}\\
D^{\bar a, \bar a'}_{s, R_t} = & \frac{\indicator{\bar A_t = \bar a_t}}{\prod_{j = 1}^t p_{A}(a_j | \parents(A_j| \bar a_{j-1}))}
    \\&\qquad\quad\prod_{j = 1}^{t-1} \frac{p_Z(Z_j | \parents(Z_j| \bar a'_j))}{p_Z(Z_j | \parents(Z_j| \bar a_j))} 
    \newbrace{Q_{s, Z_{t}}^{\bar a, \bar a'}(\bar R_t, \bar Z_{t-1}, \bar L_{t-1}) - 
    Q_{s, R_{t}}^{\bar a, \bar a'}(\bar R_{t-1}, \bar Z_{t-1}, \bar L_{t-1})} \\
D^{\bar a, \bar a'}_{s} = & D^{\bar a, \bar a'}_{s, L_0} + \sum_{t = 1}^K D^{\bar a, \bar a'}_{s, L_t} + D^{\bar a, \bar a'}_{s, Z_t} + D^{\bar a, \bar a'}_{s, R_t}. 
\end{align*}
\end{theorem}
The proof is given in Appendix \ref{sec:EIC_proof}. 

Note that here the Q functionals are defined as functions of conditional densities by iteratively carrying out all the integrals as in equation (\ref{eq:rewrite}). For example, 
\begin{align*}
    Q^{\bar a, \bar a'}_{s, L_t} (\bar R_t, \bar Z_t, \bar L_{t-1})
= & \sum_{l_t, \mathrm{Ch}(l_t) \setminus \bar a} \psi_s(\bar L_{t-1}, \bar l_t^K)
\prod_{j = t}^K
p_{L}(l_j | \bar R_t, \bar Z_t, \bar L_{t-1}, \bar a_j, \bar r_{t+1}^j, \bar z_{t+1}^j, \bar l_{t}^{j-1}) \\
& \prod_{j = t+1}^K
p_{Z}(z_j | \bar R_t, \bar Z_t, \bar L_{t-1}, \bar a'_j, \bar r_{t+1}^j, \bar z_{t+1}^{j-1}, \bar l_{t}^{j-1}) \\
& p_{R}(r_j | \bar R_t, \bar Z_t, \bar L_{t-1}, \bar a_j, \bar r_{t+1}^{j-1}, \bar z_{t+1}^{j-1}, \bar l_{t}^{j-1}). 
\end{align*}

\section{Initial Estimation of Data Likelihood}\label{sec:initial}

In this section, we briefly describe the steps and considerations for constructing an initial estimator, 
\begin{equation*}
p_n^0 = p^0_{n, L_0} \prod_{t=1}^K p^0_{n, A_t} \prod_{t=1}^K p^0_{n, R_t}p^0_{n, Z_t} p^0_{n, L_t},
\end{equation*}
for the conditional densities involved in the observed data likelihood (\ref{eq:full_lkd}) at $P_0$. 

For each binary variable $X \in \bar A \cup \bar R \cup \bar{Z} \cup \bar L$, we have $p_X(1 | \parents(X)) = \ex\newbracket{X | \parents(X)}$, and $p_X(0 | \parents(X)) = 1 - \ex\newbracket{X | \parents(X)}$. For categorical variables we define a sequence of binary dummy variables as $X^{j} = \indicator{X = j}$ or $X^{j} = \indicator{X = j, X \not\in \newbrace{1, \dots, j-1}}$, where we let $\parents(X^j) = \parents(X) \cup \newbrace{X^s: s = 1, \dots, j-1}$ then the estimation of $p_X$ can be achieved by modeling the set of conditional expectations of $\newbrace{\ex\newbracket{X^j | \parents(X^j)}: j}$. For continuous variables the modeling of conditional densities is more challenging. The modeling options include parametric assumptions, as well as discretized conditional densities which can be modeled with pooled hazard regression as specified in \cite{diaz2011super} where one specifies a model for $\ex[\indicator{X \in [\alpha_{v-1}, \alpha_v)} | A \geq \alpha_{v-1}, Pa(X)]$ over a grid $(\alpha_v : v = 0, 1, \dots)$ which spans the range of the continuous variable. Based on a library of models for all the variables, the super learner \citep{van2007super} can be applied to estimate $\ex\newbracket{X | \parents(X)}$ either as a convex ensemble of a library of candidate learners (convex learner) or as the best estimator among the candidates (discrete learner), decided by the cross-validated loss performance. Due to the finite sample oracle inequality \citep{van2003unified,van2007super}, the super learner is in general asymptotically equivalent to the oracle learner which is defined as the learner among the candidate learners that minimizes the loss under the true data-generating distribution. 

The highly adaptive lasso (HAL) \citep{benkeser2016highly,bibaut2019fast} is a nonparametric estimator with fast convergence rates that can be applied to model the conditional expectation objects $\ex\newbracket{X | \parents(X)}$. It has been shown that under the assumption that the true conditional expectation is càdlàg and bounded in sectional variation norm, the corresponding HAL estimator $\hat \ex\newbracket{X | \parents(X)}$ has the rate of convergence $\norm{\hat \ex\newbracket{X | \parents(X)} - \ex\newbracket{X | \parents(X)}}_{P_0} = O_P(n^{-1/3})$ up to $\log(n)$ factors. Therefore, any binary or categorical density estimator $p_{n, X}^0$ that is derived from a HAL (or a super learner that included HAL in the library of candidate learners) estimator of $\ex\newbracket{X | \parents(X)}$ would preserve the same convergence rate of $\norm{p_{0, X} - p_{n, X}^0}_{P_0} = O_P(n^{-1/3})$ up to $\log(n)$ factors, where $p_0 = p_{0, L_0} \prod_{t=1}^K p_{0, A_t} \prod_{t=1}^K p_{0, R_t}p_{0, Z_t} p_{0, L_t}$ is the factorized joint density at the true data distribution $P_0$. 

Lastly, we note that recent work by \texttt{haldensify} \citep{hejazi2022haldensify} allows flexible estimation of conditional densities based on HAL for discrete and continuous variables. It is recommended to consider a super learner with a library that includes HAL based estimators along with other density estimators such as kernel based or neural network based estimators. 

\section{Targeted Maximum Likelihood Estimation}\label{sec:tmle}

Given an initial density estimator $p_n^0 = p^0_{n, L_0} \prod_{t=1}^K p^0_{n, A_t} \prod_{t=1}^K p^0_{n, R_t}p^0_{n, Z_t} p^0_{n, L_t}$, we now define the TMLE updates $p_n^* = p^*_{n, L_0} \prod_{t=1}^K p^*_{n, A_t} \prod_{t=1}^K p^*_{n, R_t}p^*_{n, Z_t} p^*_{n, L_t}$ such that the plug-in estimators
\sloppy{
$\newparethensis{\Psi^{\bar a, \bar a}_s(P_n^*), \Psi^{\bar a, \bar a'}_s(P_n^*), \Psi^{\bar a', \bar a'}_s(P_n^*): s = 1, \dots, S}$ are consistent and asymptotically linear under regularity assumptions. 
}

\subsection{Iterative TMLE}\label{sec:iterative}

Define the following (locally) least favorable paths for the components of the likelihood: 
\begin{align*}
    \tilde p_{R_t}({p}_{R_t}, \bar \epsilon_{R_t}) = & (1 + \sum_{s = 1}^S \epsilon^{\bar a, \bar a}_{s, R_t} D^{\bar a, \bar a}_{s, R_t}(P) + \epsilon^{\bar a, \bar a'}_{s, R_t} D^{\bar a, \bar a'}_{s, R_t}(P) + \epsilon^{\bar a', \bar a'}_{s, R_t} D^{\bar a', \bar a'}_{s, R_t}(P)) {p}_{R_t} \\
    \tilde p_{Z_t}({p}_{Z_t}, \bar \epsilon_{Z_t}) = & (1 + \sum_{s = 1}^S 
    \epsilon^{\bar a, \bar a}_{s, Z_t} D^{\bar a, \bar a}_{s, Z_t}(P) + 
    \epsilon^{\bar a, \bar a'}_{s, Z_t} D^{\bar a, \bar a'}_{s, Z_t}(P) + 
    \epsilon^{\bar a', \bar a'}_{s, Z_t} D^{\bar a', \bar a'}_{s, Z_t}(P)
    ) {p}_{Z_t} \\
    \tilde p_{L_t}({p}_{L_t}, \bar \epsilon_{L_t}) = & (1 + \sum_{s = 1}^S 
    \epsilon^{\bar a, \bar a}_{s, L_t} D^{\bar a, \bar a'}_{s, L_t}(P) + 
    \epsilon^{\bar a, \bar a'}_{s, L_t} D^{\bar a, \bar a'}_{s, L_t}(P) + 
    \epsilon^{\bar a', \bar a'}_{s, L_t} D^{\bar a', \bar a'}_{s, L_t}(P)
    ) {p}_{L_t}. 
\end{align*}
Suppose that at each node $X \in \bar R \cup \bar Z \cup \bar L$ we have constructed a submodel of $\mathcal{M}$ with a multi-dimensional parameter $\bar \epsilon_{X} = (\epsilon^{\bar a, \bar a}_{s, X}, \epsilon^{\bar a, \bar a'}_{s, X}, \epsilon^{\bar a', \bar a'}_{s, X}: s = 1, \dots, S)$, where $\tilde p_X(p_X, \bar \epsilon_X = \bar 0) = p_X$, and the scores $\frac{\mathrm{d}}{\mathrm{d}\bar \epsilon_X}_{|\bar \epsilon_X = 0}\log \tilde p_X(p_X, \bar \epsilon_X) $ at $\bar \epsilon_X = 0$ equal the vector of the corresponding components of the efficient influence curve, $\newparethensis{D^{\bar a, \bar a}_{s, X}, D^{\bar a, \bar a'}_{s, X}, D^{\bar a', \bar a'}_{s, X}: s = 1, \dots, S}$. The maximum likelihood solutions of the parametric submodels at $p = p_n^0$ are for $t = 1, \dots, K$, 
\begin{alignat*}{2}
    \bar \epsilon_{R_t, n} 
    = & \mathrm{argmax}_{\bar \epsilon_{R_t}} {
    P_n \log \tilde { p}_{R_t}(p^0_{n, R_t}, \bar \epsilon_{R_t})
    }  \\
        \bar \epsilon_{Z_t, n} 
        = & \mathrm{argmax}_{\bar \epsilon_{Z_t}} {
    P_n \log \tilde { p}_{Z_t}(p^0_{n, Z_t}, \bar \epsilon_{Z_t})
    }  \\
        \bar \epsilon_{L_t, n} 
    = & \mathrm{argmax}_{\bar \epsilon_{L_t}} {
    P_n \log \tilde { p}_{L_t}(p^0_{n, R_t}, \bar \epsilon_{L_t})
    }, 
\end{alignat*}
which lead to the first round of TMLE updates
\begin{align*}
    \tilde p_{n, R_t} = & \tilde p_{R_t}(p_{n, R_t}^0, \bar \epsilon_{R_t, n}) \\
    \tilde p_{n, Z_t} = & \tilde p_{Z_t}(p_{n, Z_t}^0, \bar \epsilon_{Z_t, n}) \\
    \tilde p_{n, L_t} = & \tilde p_{L_t}(p_{n, L_t}^0, \bar \epsilon_{L_t, n}),
\end{align*}
and 
\sloppy{
$\tilde p_n = p^0_{n, L_0} {\prod_{t=1}^K p^0_{n, A_t} }
{\prod_{t=1}^K 
\tilde p_{n, R_t}
\tilde p_{n, Z_t}
\tilde p_{n, L_t}}$. 
}

The procedure is repeated after replacing the initial estimator $p_n^0$ with the TMLE update $\tilde p_n$ of the last iteration, till $P_n D_{s}^{\bar a, \bar a}(\tilde P_n) \to 0$, $P_n D_{s}^{\bar a, \bar a'}(\tilde P_n) \to 0$, $P_n D_{s}^{\bar a', \bar a'}(\tilde P_n) \to 0$, for $s = 1, \dots, S$. We now define the final TMLE update $p_n^*$ as the update of the $I$-th iteration, where $I=I(n)$ is large enough so that $P_n D_s^{\bar a, \bar a}(P_n^*)$, $P_n D_s^{\bar a, \bar a'}(P_n^*)$, $P_n D_s^{\bar a', \bar a'}(P_n^*)$ are all $o_P(n^{-1/2})$. The last statement is an application of the Result 1 and Theorem 1 in \cite{van2006targeted}
under 
the regularity conditions that $\tilde p_n$ is in the interior of $\mathcal{M}$ so that $P_n \frac{D^{\bar a, \bar a}_{s, X}(\tilde P_n)}{\tilde p_{n, X}} = 
P_n \frac{D^{\bar a, \bar a'}_{s, X}(\tilde P_n)}{\tilde p_{n, X}} = 
P_n \frac{D^{\bar a', \bar a'}_{s, X}(\tilde P_n)}{\tilde p_{n, X}} = 0$, $s = 1, \dots, S$, and that $\bar \epsilon_{X, n} \to 0$ for all $X \in \bar R \cup \bar Z \cup \bar L$ as the log-likelihood components $P_n \log \tilde p_{n, X}$ increase and converge with iterations. 

\begin{algorithm}
	\caption{Iterative TMLE} 
	\begin{algorithmic}[1]
	\Repeat
		\For {$X \in \bar{R}\cup\bar{Z}\cup\bar{L}$} 
		    \State Calculate clever covariates $H_{s, X}^{I} = D_{s, X}^{I}(P_n^0)$ which are equal to the EIC components for the $s$-th outcome $s=1,\dots,S$ and intervention $I = (\bar a, \bar a), (\bar a, \bar a'), (\bar a', \bar a')$. 
		    \State Find MLE of $\bar \epsilon_X = (\epsilon_{s, X}^I: s, I)$ along the multivariate submodel $(1 + \bar \epsilon^\intercal \bar H)p$ $$\bar\epsilon_{X, n} = \arg\max_{\bar \epsilon_X} P_n \log (1 + \sum_{s, I}\epsilon_{s, X}^{I}H_{s, X}^{I})p_{n, X}^0. $$
		    \State Define updates $\tilde p_{n, X}$ of $p_{n, X}^0$ by plugging $\bar \epsilon_{X, n}$ back to the submodel. 
		\EndFor
		    \State Replace $p_n^0$ with updated $\tilde p_n$. 
	\Until $P_n D_{s}^I(\tilde P_n) \leq \sqrt{\frac{\mathrm{var}_n D_{s}^I(\tilde P_n)}{n}}/\log n $ for all $s$, $I$. Let $P_n^* = \tilde P_n$. 
	\end{algorithmic} 
\end{algorithm}

\subsection{One-step TMLE}\label{sec:onestep}

We are interested to restrict the step size of $\bar \epsilon$ in each iteration of the algorithm which searches for the MLEs. This should increase the performance of the algorithm. In fact, with the Euclidian norm $\norm{\bar \epsilon} < dx$ for some small enough $dx > 0$, due to linear approximations, we have the close form MLEs
\begin{align*}
    \bar\epsilon_{X, n} = \newparethensis{P_n D_{s, X}^{\bar a, \bar a}(P_n^0), 
    P_n D_{s, X}^{\bar a, \bar a'}(P_n^0), 
    P_n D_{s, X}^{\bar a', \bar a'}(P_n^0) : s = 1, \dots, S}dx / \norm{P_n \bar D(P_n^0)},
\end{align*}
where $\bar \epsilon = \newparethensis{\bar\epsilon_X: X \in \bar R \cup \bar Z \cup \bar L}$ is the vector of all elements of $\bar \epsilon_X$'s and $\bar D = \newparethensis{D_{s, X}^{\bar a, \bar a}(P_n^0), 
    D_{s, X}^{\bar a, \bar a'}(P_n^0), 
    D_{s, X}^{\bar a', \bar a'}(P_n^0) : X \in \bar R \cup \bar Z \cup \bar L,  s = 1, \dots, S}$ is the vector of all involved EIC components (Section 8.2 of \cite{van2016one}). 

Denote by $P_{dx}$ the TMLE update of $P$ under the restriction above that $\norm{\bar \epsilon} < dx$, and by $P_{2dx}, P_{3dx}, \dots$ the TMLE updates of the next iterations. Then we can design a univariate universal least favorable path $P_{ulfm}(P, x)$ with the following analytic expression: 
\begin{align*}
    p_{ulfm}(p, x) = p\exp\newbrace{\int_0^x \frac{
    \newbrace{P_n \bar D(P_{ulfm}(p, u))}^\top \bar D(P_{ulfm}(p, u))}{\norm{P_n \bar D(P_{ulfm}(p, u))}} du}
\end{align*}
such that for all $x \in \newparethensis{0, a}$ with some $a > 0$, $\frac{d}{dx} P_n \log p_{ulfm}(p_n^0, x) = \norm{P_n \bar D(P_{ulfm}(P_n^0, x))}$. Furthermore, $\newbrace{P_{jdx}: j}$ approximates the universal path in the sense that $P_{ulfm}(P_n^0, jdx) = P_{jdx}$ for $j = 1, 2, \dots$, which can be verified under mild regularity conditions with a Taylor expansion of $p_{ulfm}(p_n^0, x)$ at $x = 0, dx, 2dx, \dots$, while the log-likelihood $P_n \log p_{j dx}$ is increasing as $j \to \infty$. 

We assume that $\newbrace{P_{ulfm}(P_n^0, x): x \in (0, a)} \subset \mathcal{M}$ is a submodel contained in $\mathcal{M}$, and that the log-likelihood $P_n \log p_{ulfm}(P_n^0, x), x \in (0, a)$ is non-decreasing along $x = dx, 2dx, \dots$ and achieves the closest local maximum at some $x_m < a$, where for small enough $dx$ and some $j_0$ we have $j_0dx = x_m$  and $\norm{P_n \bar D(P_{j_0dx})} = 0$. 
One can choose the final TMLE update to be $P_n^* = P_{Idx}$ with large enough $I = I(n)$ such that $\norm{P_n \bar D(P_{Idx})} = o_P(n^{-1/2})$. This is a one-step procedure of searching for the closest local maximum $x_m$ of $P_n \log p_{ulfm}(p_n^0, x)$ around $x=0$, where $\norm{P_n\bar D(P)}=0$ is solved exactly at $P = P_{ulfm}(P_n^0, x_m)$; hence the term ``one-step TMLE''. 

Compared to the iterative TMLE with unrestricted $\bar \epsilon$, the one-step TMLE solves the efficient score equations under weaker assumptions, 
since $P_{jdx}$'s remain in the interior of $\mathcal{M}$ so long as $P_n^0$ is an interior point of $\mathcal{M}$ and $dx$ is  chosen small enough. By conceptually searching for a local likelihood maximum of a univariate submodel, and practically updating along known directions instead of iteratively solving multi-dimensional MLEs, the one-step TMLE achieves not only greater stability but also reduced computational costs. 

\begin{algorithm}
	\caption{One-step TMLE} 
	\begin{algorithmic}[1]
	\Repeat
		\For {$X \in \bar{R}\cup\bar{Z}\cup\bar{L}$} 
		    \State Calculate clever covariates $H_{s, X}^{I} = P_n D_{s, X}^{I}(P_n^0)$ for the $s$-th outcome $s=1,\dots,S$ and intervention $I = (\bar a, \bar a), (\bar a, \bar a'), (\bar a', \bar a')$. 
	    \EndFor
	    \For {$X \in \bar{R}\cup\bar{Z}\cup\bar{L}$} 
		    \State Define known multivariate MLE of $\bar \epsilon_X = (\epsilon_{s, X}^I: s, I)$ with small enough $dx$: $$\epsilon_{s, X, n}^{I} = H_{s, X}^I dx/\norm{\left(\sum_X H_{s, X}^I: s, I\right)}  $$
		    \State Define updates $\tilde p_{n, X}$ of $p_{n, X}^0$ by plugging $\bar \epsilon_{X, n}$ back to the submodel. 
		\EndFor
		    \State Replace $p_n^0$ with updated $\tilde p_n$. 
	\Until $P_n D_{s}^I(\tilde P_n) \leq \sqrt{\frac{\mathrm{var}_n D_{s}^I(\tilde P_n)}{n}}/\log n $ for all $s$, $I$. Let $P_n^* = \tilde P_n$. 
	\end{algorithmic} 
\end{algorithm}

\subsection{Asymptotic Efficiency and Simultaneous Inference} \label{sec:efficiency}

For the purpose of longitudinal mediation analysis, suppose that the target parameter is a vector $\bar \Psi^F(P)$ of a differentiable transformations of $\newparethensis{\Psi^{\bar a, \bar a}_s(P), \Psi^{\bar a, \bar a'}_s(P), \Psi^{\bar a', \bar a'}_s(P): s = 1, \dots, S}$, where the EIC vector $\bar D^F(P)$ is calculated as functions of $\newparethensis{D^{\bar a, \bar a}_s(P), D^{\bar a, \bar a'}_s(P), D^{\bar a', \bar a'}_s(P): s = 1, \dots, S}$ according to the functional delta method (see A.3 of \cite{van2011targeted} and Section 3 of \cite{van2004asymptotic}). In this section we consider $P_n^*$ to be an iterative or one-step TMLE, as described in Section \ref{sec:iterative} and Section \ref{sec:onestep}, which under regularity conditions achieves $P_n D_s^{\bar a, \bar a}(P_n^*) = o_P(n^{-1/2})$, $P_n D_s^{\bar a, \bar a'}(P_n^*) = o_P(n^{-1/2})$, $P_n D_s^{\bar a', \bar a'}(P_n^*) = o_P(n^{-1/2})$ for $s = 1, \dots, S$. 

We define the remainder for the target parameter as 
\begin{align*}
\bar R^{F}(P, P_0) = \bar \Psi^F(P) - \bar \Psi^F(P_0) + P_0 \bar D^F(P).
\end{align*}
The remainder for the target parameters are functions of the remainders $\newparethensis{R^{\bar a, \bar a}_s(P, P_0), R^{\bar a, \bar a'}_s(P, P_0), R^{\bar a', \bar a'}_s(P, P_0): s = 1, \dots, S}$ for each of the corresponding targets in $\newparethensis{\Psi^{\bar a, \bar a}_s(P), \Psi^{\bar a, \bar a'}_s(P), \Psi^{\bar a', \bar a'}_s(P): s = 1, \dots, S}$. 
The TMLE $P_n^*$ satisfies the following expansion: 
\begin{align*}
     \bar \Psi^F(P_n^*) - \bar \Psi^F(P_0) = & \bar R^{F}(P_n^*, P_0) - P_0 \bar D^F(P_n^*) \\
    = & (P_n - P_0) \bar D^F(P_1) + (P_n - P_0) (\bar D^F(P_n^*) - \bar D^F(P_1)) 
    \\ & - P_n \bar D^F(P_n^*) + \bar R^{F}(P_n^*, P_0). \numberthis \label{eq:exact_expansion}
\end{align*}
Asymptotic linearity of $P_n^*$ is achieved for $\bar \Psi^F(P)$ under the following conditions: 
\begin{enumerate}[label=(B\arabic*)]
    \item\label{cond:donsker} The vector of EIC $\newparethensis{D^{\bar a, \bar a}_s(P), D^{\bar a, \bar a'}_s(P), D^{\bar a', \bar a'}_s(P): s = 1, \dots, S}$ at $P = P_n^*$ converges to its limit at $P = P_1$ in $L^2(P_0)$ norm on each dimension, and falls in a $P_0$-Donsker class; 
    \item Elements of $\newparethensis{R^{\bar a, \bar a}_s(P_n^0, P_0), R^{\bar a, \bar a'}_s(P_n^0, P_0), R^{\bar a', \bar a'}_s(P_n^0, P_0): s = 1, \dots, S}$ are $o_P(n^{-1/2})$. \label{assumption_res}
\end{enumerate}
Asymptotic efficiency is achieved under the following additional condition: 
\begin{enumerate}[resume,label=(B\arabic*)]
    \item The limit in Assumption \ref{cond:donsker} is achieved at $P_1 = P_0$. 
\end{enumerate}

We can then construct simultaneous confidence intervals  based on the asymptotic linearity and the normal limit distribution \citep{dudoit2008multiple,rose2018ltmle}. Note that under the above conditions, the following empirical covariance matrix, 
\begin{align*}
    \bar \Sigma(P_n, P_n^*) = P_n \bar D^F(P_n^*) \bar D^F(P_n^*)^\top, 
\end{align*}
provides consistent estimation of the limit covariance matrix $\bar \Sigma(P_0, P_0)$. Therefore, the $95\%$ simultaneous confidence interval can be constructed as 
\begin{align*}
    \bar\Psi^F(P_n^*) \pm q_{0.95, n} \bar \sigma_n / \sqrt{n}, 
\end{align*}
where $\bar \sigma_n$ is the vector of diagonal elements of $\newbrace{\mathrm{diag}\newparethensis{\bar \Sigma(P_n, P_n^*)}}^{1/2}$, and $q_{0.95, n}$ can be a Monte-Carlo estimate of the $0.95$ quantile of the maximum of element-wise absolute values of a random vector $\bar Z$ that follows multivariate normal with mean $\bar 0$ and covariance matrix $\newbrace{\mathrm{diag}\newparethensis{\bar \Sigma(P_n, P_n^*)}}^{-1/2}  \bar \Sigma(P_n, P_n^*)  \newbrace{\mathrm{diag}\newparethensis{\bar \Sigma(P_n, P_n^*)}}^{-1/2}$.

\section{Numerical Improvements}\label{sec:num}

One computational challenge of conducting TMLE updates for all factors of the likelihood simultaneously lies in the need of repeatedly calculating nested integrals with respect to conditional densities of the kind shown in equation (\ref{eq:rewrite}). Here we derive an alternative projection representation of EIC, which is called HAL-EIC, and then show that a numerical approximation of HAL-EIC can reduce the computational costs while preserving the asymptotic properties under mild conditions. Throughout this section, for notation simplicity we focus on the EIC $D^{\bar a, \bar a'}$ of a single real valued target parameter. But the methods also apply to multidimensional target parameters in our setting. 

\subsection{Numerical Approximation of the EIC based on HAL-EIC} \label{sec:HAL_EIC}

For any variable $X \in \bar R \cup \bar Z \cup \bar L$ let $T_X(P) = \newbrace{f \in L^2(P): \ex\newbracket{f | \parents(X)} = 0}$ denote the tangent subspace, and for any $f \in L^2(P)$ define the projection onto \(T_X(P)\) with respect to $L^2(P)$ norm as $\prod(f | T_X(P)) = \arg\min_{h \in T_X(P)} P\newbrace{f - h}^2$.
\begin{lemma}[Projection representation of EIC]\label{lemma:projection}
Define initial mappings as 
\begin{align*}
    G^{\bar a, \bar a'}_{LR} = & G^{\bar a, \bar a'}_{L} = G^{\bar a, \bar a'}_{R} = \frac{Y\indicator{\bar{A} = \bar{a}}}{\prod_{j=1}^K p_A(a_j | \parents(A_j | \bar{a}_{j-1}))} \frac{\prod_{j=1}^K p_Z(Z_j | \parents(Z_j | \bar{a}'_{j}))}{\prod_{j=1}^K p_Z(Z_j | \parents(Z_j | \bar{a}_{j}))} \\
    G^{\bar a, \bar a'}_{Z} = & \frac{Y \indicator{\bar{A} = \bar{a'}}}{\prod_{j=1}^K p_A(a'_j | \parents(A_j | \bar{a}'_{j-1}))} \frac{\prod_{j=1}^K p_L(L_j | \parents(L_j | \bar{a}_{j}))}{\prod_{j=1}^K p_L(L_j | \parents(L_j |\bar{a}'_{j}))}
    \frac{\prod_{j=1}^K p_R(R_j | \parents(R_j | \bar{a}_{j}))}{\prod_{j=1}^K p_R(R_j | \parents(R_j | \bar{a}'_{j}))}.  
\end{align*}
The following projection representation holds for $t = 1, \dots, K$: 
{
    \begin{align*}
    D^{\bar a, \bar a'}_{L_t}(P) = & \prod(G_{L}^{\bar a, \bar a'}(P) | T_{L_t}(P)) \\
    D^{\bar a, \bar a'}_{R_t}(P) = & \prod(G_{R}^{\bar a, \bar a'}(P) | T_{R_t}(P)) \\
    D^{\bar a, \bar a'}_{Z_t}(P) = & \prod(G_{Z}^{\bar a, \bar a'}(P) | T_{Z_t}(P)). 
    \end{align*}
}
\end{lemma}

The proof is given in Appendix \ref{sec:projection_proof}. 

Note that the projection terms in Lemma \ref{lemma:projection} can be considered as true risk minimizers in tangent subspaces, if $P$ is considered as the ``true'' distribution and the risk $\norm{G^{\bar a, \bar a'}_X - f}_{P}^2 = P\newbrace{G^{\bar a, \bar a'}_X - f}^2$ is defined for all $f \in T_X(P) \subset L^2(P)$. Given an IID sample following $P$, the approximation of the projection terms can be done by empirical risk minimizers over a class of functions that contains the true EIC. Here we adopt an additional regularity condition in order to introduce the HAL approximation of the EIC and to achieve fast convergence rates \citep{bibaut2019fast}:
\begin{enumerate}[resume,label=(B\arabic*)]
    \item For $X \in \bar R, \bar Z, \bar L$, the corresponding EIC components at the true distribution $P_0$, the initial estimator $P_n^0$, and at any TMLE update $\tilde P_n$ are cadlag with bounded sectional variation norm. \label{assumption_bound}
\end{enumerate}
Under condition \ref{assumption_bound}, we construct the centered HAL basis as 
$$\phi_{j, X}(P) = \indicator{\parents(X) \geq \parents(x)(\mu_j)}\newparethensis{
\indicator{X \geq x(\mu_j)} - P(X \geq x(\mu_j) | \parents(X))
},$$ 
where $\newbrace{u_{j}: j}$ is the set of knot points on $(X, \parents(X))$, and $u_j = \newparethensis{x(u_{j}), \parents(x)(u_{j})}$ are the corresponding subvectors. Note that these centered HAL bases satisfy (as we show in Appendix \ref{sec:basis}): 1) $T_{X}(P)$ is spanned by the collection $\newbrace{\phi_{j, X}(P): j}$, and 2) given an IID sample of size $N$ following $P$, the lasso regression of $G^{\bar a, \bar a'}_{X}(P)$ uses a finite subset of $\newbrace{\phi_{j, X}(P): j}$ with a bound on sectional variation norms decided by cross-validation can achieve a guaranteed convergence rate, where the lasso estimator
\begin{align*}
    \tilde D^{\bar a, \bar a'}_X(P) = \sum_{j=1}^J \hat\beta^{\bar a, \bar a'}_{j, X}(P) \phi_{j, X}(P)
\end{align*} 
approximates the EIC with $\norm{\tilde D^{\bar a, \bar a'}_{X}(P) - D^{\bar a, \bar a'}_X(P)}_{P} = O_P(N^{-1/3}(\log N)^{d/2})$. 
We call the numerical approximation 
$$\tilde D^{\bar a, \bar a'}(P) = \sum_{X \in \bar R \cup \bar Z\cup\bar L}\tilde D^{\bar a, \bar a'}_X(P)$$ 
 the HAL-EIC for $D^{\bar a, \bar a'}(P)$. In what follows we refer to the iterative and the one-step TMLE which replaces the EIC with the HAL-EIC as iterative and one-step HAL-EIC TMLE, respectively. 

Note that the approximation $\tilde D^{\bar a, \bar a'}_{X}(\cdot) \approx D^{\bar a, \bar a'}_X(\cdot
)$ can be obtained for the initial estimate $P_n^0$ or for a TMLE update $\tilde P_n$. But the coefficients $\hat \beta^{\bar a, \bar a'}_{j, X}(P)$ of the LASSO estimator are functions of $P$, and thus the estimation of the $\beta$ coefficients requires an IID sample of $N$ random vectors with joint distribution $P$. The resampling size $N$ can be chosen to be larger than the observed sample size $n$. 

To achieve computationally fast HAL-EIC TMLE updates, we note 
that data resampling of size $N$ 
needs not happen for all $P$ in real time as $P$ changes. Instead, we define HAL-EIC with delayed coefficient estimation by
\begin{align*}
    \tilde D^{\bar a, \bar a'}_{X, P_n^0}(P) = \sum_{j=1}^J \hat \beta^{\bar a, \bar a'}_{j, X}(P_n^0) \phi_{j, X}(P), 
\end{align*} 
and 
$$\tilde D^{\bar a, \bar a'}_{P_n^0}(P) = \sum_{X \in \bar R \cup \bar Z\cup\bar L}\tilde D^{\bar a, \bar a'}_{X, P_n^0}(P), $$ 
where we keep the $\beta$ coefficients unchanged with respect to a given initial estimate $P_n^0$. Resampling and re-estimation ofthese $\beta$ coefficients will only happen when the value of $P_n^0$ is changed. 
For the fast version of HAL-EIC TMLE, define $\tilde P_n$ as the (iterative or one-step) TMLE update of $P_n^0$ that solves $P_n \tilde D^{\bar a, \bar a'}_{X, P_n^0}(\tilde P_n) = o_P(n^{-1/2})$. Repeat the procedure for $I$ iterations by replacing $P_n^0$ with $\tilde P_n$ at the end of each iteration. Then, under the same regularity conditions of iterative or one-step TMLE, there exists a large enough integer $I = I(n)$ such that at the $I$-th iteration we have $P_n \tilde D^{\bar a, \bar a'}_{X}(\tilde P_n) = P_n \tilde D^{\bar a, \bar a'}_{X, \tilde P_n}(\tilde P_n) = o_P(n^{-1/2})$, and we define the final TMLE update $P_n^*$ as the TMLE update $\tilde P_n$ of the $I$-th iteration.

\begin{algorithm}
	\caption{HAL-EIC (One-step) TMLE. } 
	\begin{algorithmic}[1]
	
	    		\For {$X \in \bar{R}\cup\bar{Z}\cup\bar{L}$} 
    		    \State Initialize numerical HAL-EIC $\tilde D_{s, X, P_{\text{fixed}}}^{I}(P_n^0)$ for $X \in \bar{R}\cup\bar{Z}\cup\bar{L}$ with fixed $\beta$ coefficients fitted at generated IID samples from $P_{\text{fixed}} = P_n^0$, for the $s$-th outcome $s=1,\dots,S$ and intervention $I = (\bar a, \bar a), (\bar a, \bar a'), (\bar a', \bar a')$. Calculate clever covariates $H_{s, X}^{I}(P_n^0) = P_n \tilde D_{s, X, P_{\text{fixed}}}^{I}(P_n^0)$. 
    		\EndFor

    \Repeat 

	\Repeat
		
	    \For {$X \in \bar{R}\cup\bar{Z}\cup\bar{L}$} 
		    \State Define known multivariate MLE of $\bar \epsilon_X = (\epsilon_{s, X}^I: s, I)$ with small enough $dx$: $$\epsilon_{s, X, n}^{I} = H_{s, X}^I(P_n^0) dx/\norm{\left(\sum_X H_{s, X}^I(P_n^0): s, I\right)}  $$
		    \State Define updates $\tilde p_{n, X}$ of $p_{n, X}^0$ by plugging $\bar \epsilon_{X, n}$ back to the submodel. 
		\EndFor
		\State    Replace $p_n^0$ with updated $\tilde p_n$. 
	\Until $P_n \tilde D_{s, P_{\text{fixed}}}^{I}(\tilde P_n) \leq \sqrt{\frac{\mathrm{var}_n \tilde D_{s, P_{\text{fixed}}}^{I}(\tilde P_n)}{n}}/\log n $ for all $s$, $I$.
	\State Refit $\beta$ coefficients in the numerical EIC objects $\tilde D_{s, X, P_{\text{fixed}}}^{I}(P)$ and also update $H_{s, X}^{I}(P_n^0) = \tilde D_{s, X, P_{\text{fixed}}}^{I}(P_n^0)$ by replacing $P_{\text{fixed}} = P_n^0$ where $P_n^0$ has just been replaced with the most recent TMLE update $\tilde P_n$. 
	\Until $P_n \tilde D_{s, P_{\text{fixed}}}^{I}(\tilde P_n) \leq \sqrt{\frac{\mathrm{var}_n \tilde D_{s, P_{\text{fixed}}}^{I}(\tilde P_n)}{n}}/\log n $ for all $s$, $I$ with updated $\beta$ coefficients. 
	\end{algorithmic} 
\end{algorithm}

The asymptotic linearity and efficiency of HAL-EIC TMLE are preserved under the following additional conditions on the resampling size $N$ for the lasso estimators of the $\beta$ coefficients (see Appendix \ref{sec:hal_eic_tmle_proof}): 
\begin{enumerate}[resume,label=(B\arabic*)]
    \item $N=n$ and $P_0\newbrace{p_0 - \tilde{p}_n}^2 = o_P(n^{-1/3})$, or $N(n)$ increases faster than $n^{3/2}$, under the strong positivity condition that $p_0(o) > \delta > 0$ and $p_n^*(o) > \delta > 0$ over the supports for some $\delta > 0$. \label{assumption_two}
\end{enumerate}

\subsection{Modeling with Summary Covariates}\label{sec:dimr}

Due to the curse of dimensionality and challenges of modeling conditional densities, it is of interest in practice to consider dimension reductions by introducing summary covariates. For example, for each node $X \in \newbrace{\bar{A}, \bar{R}, \bar{Z}, \bar{L}}$, suppose that there exists a vector-valued deterministic function $C_{X}: \parents(X) \mapsto C_{X}(\parents(X))$ that summarizes the information in $\parents(X)$ hopefully without loosing information such that for all $P \in \mathcal{M}$:
\begin{align}
    p_X\newparethensis{X | \parents(X)} = p_X\newparethensis{X | C_X(\parents(X))}. \label{eq:reduced_model}
\end{align}
Equation (\ref{eq:reduced_model}) can be considered as enforcing an extra restriction on the statistical model $\mathcal{M}$, where the number of independent variables in each conditional density can now be decided by the dimensionality of the summary vector $C_X$, not necessarily increasing with the number of time points.

The summary covariates $C_X$ may be chosen in a data adaptive manner so that an analysis under assumption (\ref{eq:reduced_model}) can be conducted. For example, note that for a discretized categorical variable $X$ with possible levels $1, \dots, d_X$, a natural oracle choice of $C_X$ is
\begin{align*}
    C_X(\parents(X)) = \newparethensis{P(X = x | \parents(X)): x  \in \newbrace{1, \dots, d_X}}, 
\end{align*}
which satisfies (\ref{eq:reduced_model}) by iterated expectations. Intuitively, this observation is related to propensity score matching or covariate adjustment \citep{rosenbaum1983central,d1998propensity}. Although in practice such $C_X$ is not observed, and using estimated conditional probabilities might as well introduce bias, it is possible to augment $C_X$ with additional terms while utilizing HAL in modeling the conditional densities as functions of $C_X$, so that the desired asymptotic properties can be preserved. We refer readers to the technical report of meta-HAL super-learners for theoretical details. 
Under mild conditions the summary covariates $C_X$ can be obtained from training samples, and the resulting CV-TMLE \citep{zheng2011cross,hubbard2018data} will still be locally efficient for the target parameter of interest. 

In practice, the conditional density models may as well be achieved by applying actual knowledge of the data generating process. For example, if it is known that the propensity of prescribing a medicine only depends on recent onsets of a specific symptom and pre-existing conditions at the time of the prescription, then in this case $C_{A_t}$ is taking a subset of the vector $\parents(A_t)$. It may be advisable to replace $\parents(A_t)$ with its interaction set with the most recent time points prior to $t$. Then the same expressions of g-computation formulas, EIC, and tangent subspaces hold, and algorithms in the previous sections apply.

\section{Multiple Robustness}\label{sec:MR}

Multiple robustness \citep{diaz2017doubly, luedtke2017sequential} of the proposed estimators is obtained if two out of the following three sets of conditional density estimators: 1) $p^0_{n, A_t}$, 2) $p^0_{n, Z_t}$, 3) $p^0_{n, R_t}$ and $p^0_{n, L_t}$, are correct or at least consistent with $o_P(n^{-1/4})$ error rates in $L^2(P_0)$ norms. 
Then the remainders defined in $\newparethensis{R^{\bar a, \bar a}_s(P, P_0), R^{\bar a, \bar a'}_s(P, P_0), R^{\bar a', \bar a'}_s(P, P_0): s = 1, \dots, S}$ are all $o_P(n^{-1/2})$ (see Appendix \ref{sec:res_proof}). 
However, in practice the multiple robustness conditions for mediation analysis are not trivially satisfied even with randomized control trials, where only $p^0_{n, A_t}$'s are guaranteed to be correct or consistently estimated. 
Therefore, it is recommended to include HAL as one of the estimators in the super learner for $P_n^0$, so that the error rate conditions are all satisfied \citep{bibaut2019fast,van2007super,van2004asymptotic}. 

As a variant of the current setting, if one were to define a data adaptive framework so that the mediator random intervention $\Gamma_t^{\bar a'}(z_t | \bar r_t, \bar z_{t-1}, \bar l_{t-1}) = p(Z_t(\bar a') = z_t | \bar R_{t}(\bar a') = \bar r_t, \bar Z_{t-1}(\bar a') = \bar z_{t-1}, \bar L_{t-1}(\bar a') = \bar l_{t-1}) = p_{Z_t}(z_t | \parents(z_t|\bar a_t') )$ is replaced by, for example, an estimated control group mediator distribution $\Gamma^{\bar a'}_{n, t}(z_t | \bar r_t, \bar z_{t-1}, \bar l_{t-1}) = p^0_{n, Z_t}(z_t | \parents(z_t|\bar a_t'))$, then the counterfactuals $X(\bar a, \bar \Gamma_n^{\bar a'})$ and resulting target parameters would also become data adaptive \citep{hubbard2018data}.
If it is further assumed that $\Gamma_t^{\bar a'} = \Gamma_{n, t}^{\bar a'} $, then it is a generalization of \cite{van2008direct} to longitudinal data. 
In those generalizations, the multiple robustness conditions may be reliably satisfied in randomized trials with known treatment randomization and dynamic rules, but different interpretation follows for the new targets of inference, which now depends on the choice of mediator interventions. The influence curves and implementation for such generalized stochastic direct and indirect effects are discussed in Appendix \ref{sec:CDE_EIC}. 

\section{Simulations}

In this section, we investigate the properties of the proposed algorithms in simulated data. Throughout this section, we focus on the following data structure, 
\begin{align*}
    O = (L_{01}, L_{02}, A^C_1, A^E_1, R_1, Z_1, Y_1, A^C_2, A^E_2, R_2, Z_2, Y_2) \sim P_0. 
\end{align*}
We focus on the survival outcome $\bar Y = (Y_1, Y_2)$ where \(Y_1\) and \(Y_2\) are binary such that the event \(\{Y_1=1\}\) implies \(\{Y_2=1\}\). We target the multivariate parameter $\newparethensis{\ex\newbracket{\bar Y(\bar 1, \Gamma^{\bar 1})}, \ex\newbracket{\bar Y(\bar 1, \Gamma^{\bar 0})}, \ex\newbracket{\bar Y(\bar 0, \Gamma^{\bar 0})}}$. 
We calculate the TMLE (one-step TMLE with restricted step sizes) using both the exact EIC and the HAL-EIC, and the g-formula plug-in with different correct or misspecified initial estimators. This subsection aims to verify the consistency, asymptotic linearity, and  multiple robustness properties for both exact EIC or HAL-EIC based TMLE. 

Within each iteration, we generate an IID sample of size $n=1000$  w.r.t. the following data generating process:
\begin{align*}
    L_{01} \sim & \text{Bernoulli}(0.4)\\
    L_{02} \sim & \text{Bernoulli}(0.6) \\
    A^C_t \sim & \text{Bernoulli}(\text{expit}(1.5 - 0.4 L_{01} - 0.8 L_{02} + \indicator{t>1} 0.5 A^E_{t-1} )) \\
    A^E_t \sim & \text{Bernoulli}(\text{expit}(\lambda(-0.55 + 0.35 L_{01} + 0.6 L_{02} - \indicator{t>1}0.05 A^E_{t-1}) )) \\ 
    R_t \sim & \text{Bernoulli}(\text{expit}(-0.8 + 0.1 L_{01} + \indicator{t = 1}0.3 L_{02} + \indicator{t>1}0.3R_{t-1} + A^E_t )) \\ 
    Z_t \sim & \text{Bernoulli}(\text{expit}(-0.25 + 0.4 L_{02} + 0.4 A^E_t + 0.5 R_t )) \\
    Y_t \sim & \text{Bernoulli}(\text{expit}(0.05 + 0.375 L_{02} + 0.25 R_t - 0.075 A^E_t - 0.075 Z_t - \indicator{t>1}0.025 R_{t-1}  )) , 
\end{align*}
where we also vary the value of the propensity scaling factor $\lambda$ from $1$ to $5$ in order to simulate different degrees of finite-sample near-violation of the positivity assumptions. Each scenario iterates for $1000$ times and detailed results are reported in Appendix \ref{appendix:sim}. 

\subsection{Multiple Robustness}

We present results of simulation study which is a proof-of-concept study for the basic multiple robustness of exact-EIC TMLE and the comparable performance of HAL-EIC TMLE. Model misspecification was enforced to: 1) none of the conditional density estimators; 2) initial $p_A$ estimators; 3) initial $p_Z$ estimators; 4) initial $p_Y$ estimators. Correct conditional density models were fitted with correct main-term logistic regressions. Misspecified models set conditional expectations of each of the variables given the past as observed sample means with an additional bias of $0.05$ while bounded between $(0.01, 0.99)$. Exact-EIC and HAL-EIC based TMLE achieved similar performance in all four scenarios (see Table \ref{tab:none}-\ref{tab:Y}). 

\subsection{Finite Sample Positivity Challenge}

As we are pushing the positivity parameter $\lambda$ from $1$ to $5$, the true treatment propensity score gets closer to $0$, and therefore in finite samples we see that the product of the inverse estimated propensity scores at the initial estimate $p_n^0$ and the follow-up updates $\tilde p_n$ violate the boundedness assumptions \ref{assumption_res}, \ref{assumption_bound}, and \ref{assumption_two}. 

\begin{figure}
    \centering
    \includegraphics[scale = 0.72]{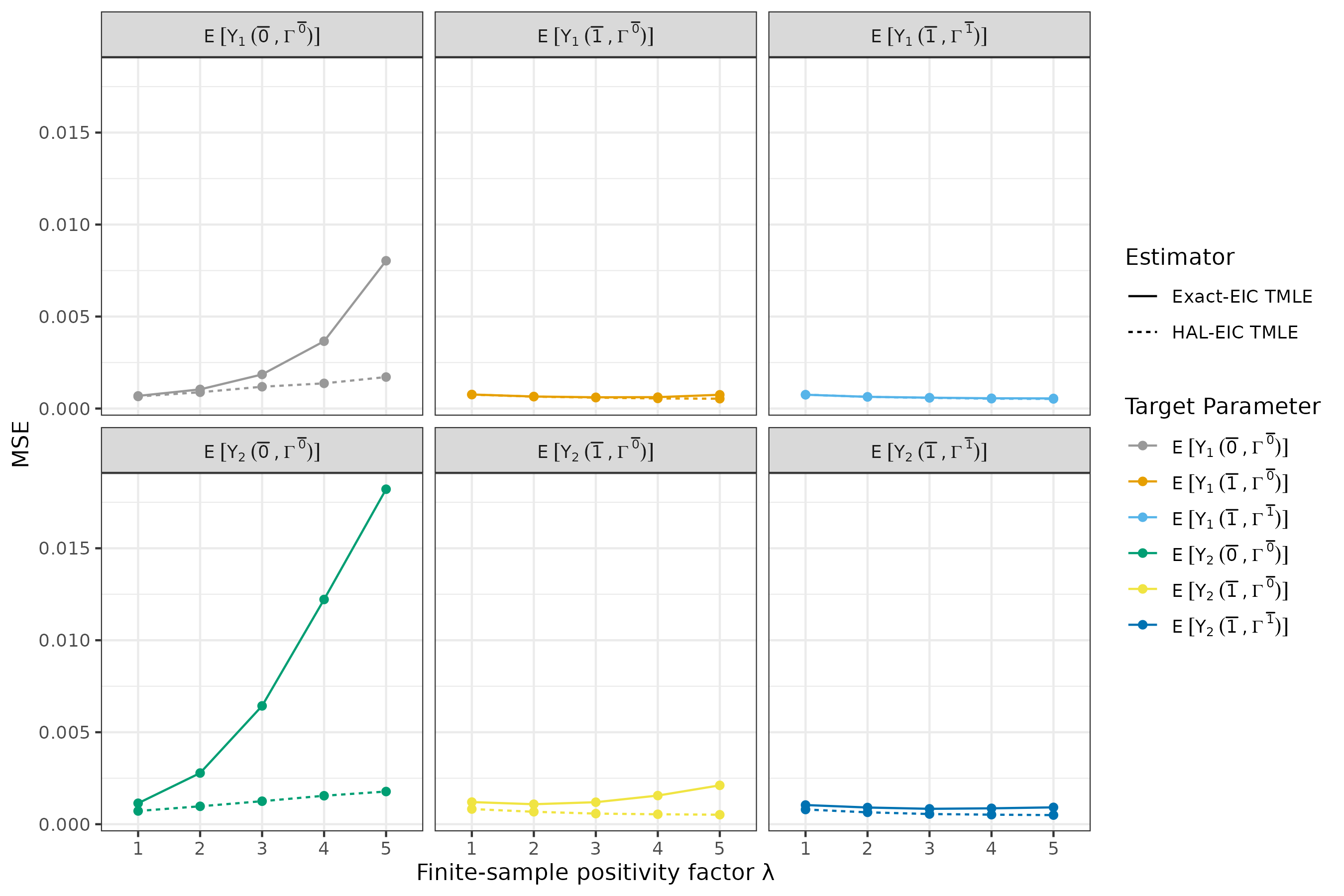}
    \caption{MSE comparison with increasing finite-sample positivity violation. }
    \label{fig:mse}
\end{figure}

\begin{figure}
    \centering
    \includegraphics[scale = 0.72]{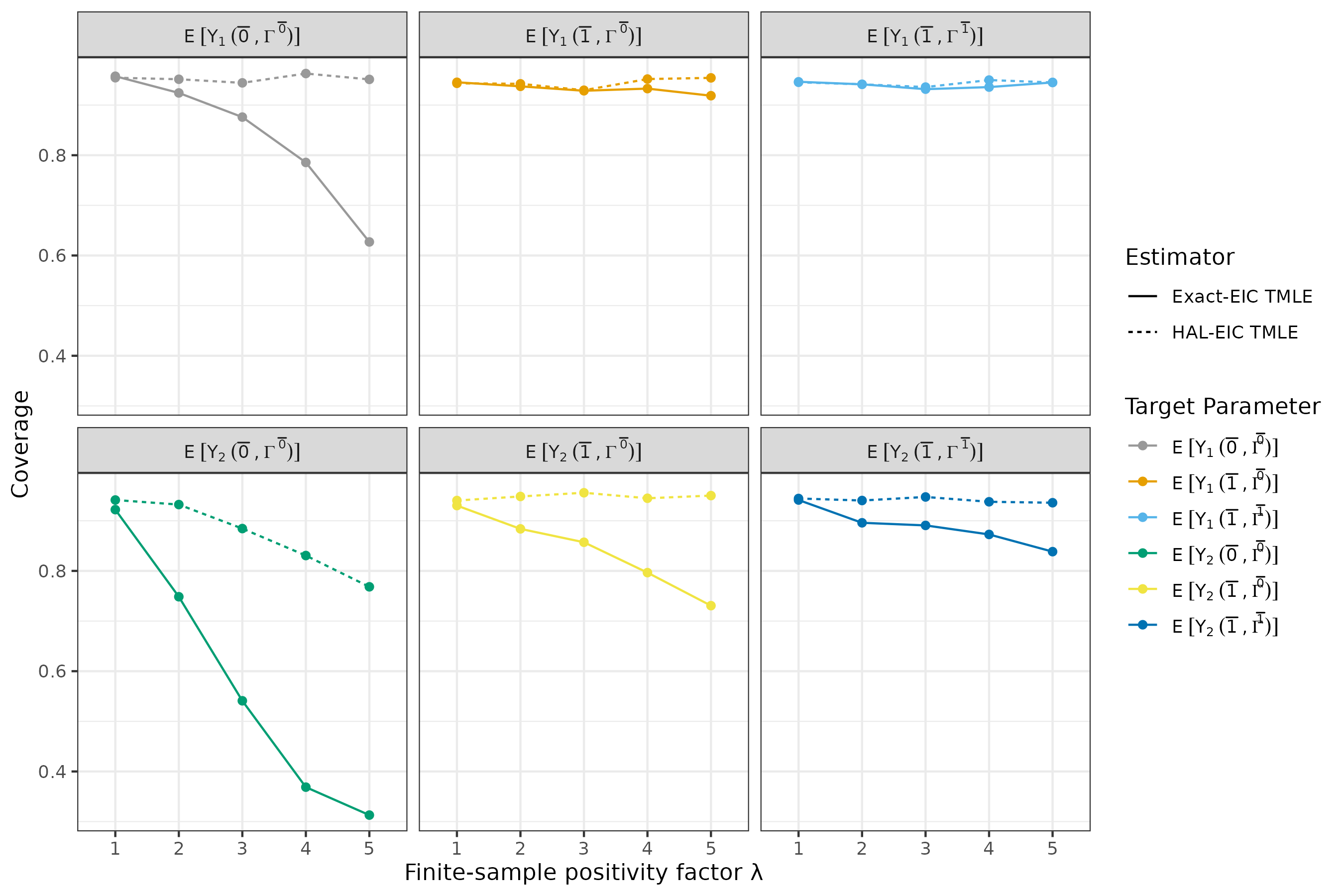}
    \caption{Coverage comparison with increasing finite-sample positivity violation.}
    \label{fig:coverage}
\end{figure}

Interestingly, in all scenarios ($\lambda = 1, 2, 3, 4, 5$) for all dimensions of the target parameter, HAL-EIC TMLE had less increase in MSE and less drop in confidence interval coverage compared to exact-EIC TMLE (Figure \ref{fig:mse}-\ref{fig:coverage}). This illustrates one potential advantage of the HAL-based projection representation: the HAL algorithm automatically searches for a bounded EIC approximation while maintaining the asymptotic linearity of the final TMLE estimate. This is a desirable property and more appealing than to arbitrarily set bounds on the IPW value or on the influence curve estimates. The latter tend to create asymptotic bias, while the HAL-EIC maintains desired multiple robustness. 

Further investigation is needed to study the performance of the HAL-EIC in more complex scenarios such as rare events in combination of finite-sample positivity violation. This and further numerical optimizations will be addressed in future venue. 

\section{Discussion}

In this manuscript, we construct a scalable, likelihood and random intervention based estimation framework for longitudinal mediation. One advantage of our framework is its flexibility being a likelihood based method. This leads to an estimation procedure that is able to target multi-dimensional parameters simultaneously by updating estimators of the same estimate of the longitudinal likelihood process. Moreover, our HAL-EIC approximation further reduces the burden on practitioners, as the required input of the algorithms is reduced to modeling constraints of conditional distributions which is a well-known task for data scientists and can be implemented without further understanding of either the analytic calculation of the EIC or the sequential regression equations. 

An important extension of our work is the application of collaborative TMLE \citep{van2010collaborative}, which can be crucial in longitudinal problems where finite-sample positivity violations occur \citep{petersen2011positivity}. While the numerical HAL-EIC representation proposed in Section \ref{sec:HAL_EIC} implicitly searches for an optimal bound of the sectional variation norms on the EIC components and demonstrates robustness in the simulation, collaborative TMLE may be combined with the proposed estimators to achieve even more reliable inference. Previous results show that this is a promising direction to optimize the finite-sample performance \citep{ju2019adaptive} . 

Another idea that needs further work is a data adaptive choice of a dimension reduction algorithm for applications with high dimensional time-varying covariates. 
This can be conducted so that scalability is further improved  while maintaining asymptotic properties for the estimators of the target parameters. 

Future research should also consider generalize to continuous time TMLE \citep{rytgaard2021continuous} for applications where it is of interest to apply longitudinal mediation analysis to  data structures where events are happening and are recorded in continuous time. Higher order TMLE \citep{van2021higher} should be utilized to improve efficiency.

\begin{acks}[Acknowledgments]
This research is partially funded by NIH-grant R01AI074345-10A1. Additional funding was
provided by a philanthropic gift from Novo Nordisk to the Center for
Targeted Machine Learning and Causal Inference at UC Berkeley.
\end{acks}

\clearpage
\appendix

\section{Efficient Influence Curves (EIC)}\label{sec:EIC_proof}

To prove the EIC representation in Section \ref{sec:EIC}, we utilize the pathwise differentiability of $\Psi^{\bar a, \bar a'}_s(P)$ and have that (suppressing its dependence on $\bar a, \bar a', s$, and let $Y = \psi_s(\bar L_K)$)
\begin{align*}
    \frac{\mathrm{d}\Psi(P_\epsilon^h)}{\mathrm{d}\epsilon}_{|\epsilon = 0} = \langle D(P), h \rangle_P
\end{align*}
for all submodel $P^h_\epsilon$ through $P$, such that $h = \frac{\mathrm{d}}{\mathrm{d}\epsilon}_{|\epsilon = 0} \log p^h_{\epsilon}$. Note that $h \in T(P)$ can be decomposed as $h = \sum_{X} h_X$ where $X \in \bar L \cup \bar R \cup \bar Z$, which corresponds to the factorization of the submodel
\begin{align*}
    p^h_\epsilon = \prod_{X \in \bar L \cup \bar R \cup \bar Z} p^h_{\epsilon, X} = \prod_{X \in \bar L \cup \bar R \cup \bar Z} p_X(1 + \epsilon h_X). 
\end{align*}

For all $h \in T(P)$ such that $h_{Z_t} = 0$ for all $t = 1, \dots, K$, we have that 
\begin{align*}
    \frac{\mathrm{d}\Psi(P_\epsilon^h)}{\mathrm{d}\epsilon}_{|\epsilon = 0} = & 
    \frac{\mathrm{d}\Psi(
    \prod_{t = 1}^K p_{Z_t}
    \prod_{X \in \bar L \cup \bar R} p^h_{\epsilon, X}
    )}{\mathrm{d}\epsilon}_{|\epsilon = 0}\\
    = & \frac{\mathrm{d}
    \sum_{\bar l, \bar z, \bar r} y \prod_{X \in \bar L \cup \bar R} p^{h_X}_{\epsilon, X}(x | \parents(x | \bar a))
    {
    \color{red}
            \prod_{t = 1}^K\indicator{ A_t= a_t}
        \prod_{t = 1}^K p_{Z_t}(z_t | \parents(z_t | \bar a_t'))
    }
    }{\mathrm{d}\epsilon}_{|\epsilon = 0} \\
    = & \langle \sum_{X \in \bar L \cup \bar R} D_{X}(P), \sum_{X \in \bar L \cup \bar R} h_X\rangle_P
\end{align*}
which corresponds to the pathwise derivative of the treatment specific mean under interventions $A_t = a_t$ and $Z_t \sim p_{Z_t}(Z_t | \parents(Z_t | \bar a'))$. Apply $\Gamma_t(Z_t | \parents(Z_t)) = p_{Z_t}(Z_t | \parents(Z_t | \bar a'))$ in Lemma \ref{lemma:EIC_general}. This gives the efficient influence function $\sum_{X \in \bar L \cup \bar R} D_X(P)$ as specified in Section \ref{sec:EIC}. 

On the other hand, to calculate $D_{Z_t}(P)$, consider $h \in T(P)$ such that $h_X = 0$ for all $X \in \bar L \cup \bar R$. Then 
\begin{align*}
        \frac{\mathrm{d}\Psi(P_\epsilon^h)}{\mathrm{d}\epsilon}_{|\epsilon = 0} = & 
        \frac{\mathrm{d}
    \sum_{\bar l, \bar z, \bar r} y 
    {\color{red}
        \prod_{t = 1}^K\indicator{ A_t= a_t'}
        \prod_{X \in \bar L \cup \bar R} p_{X}(x | \parents(x | \bar a))
    }
    \prod_{t = 1}^K p^{h_{Z_t}}_{\epsilon, Z_t}(z_t | \parents(z_t | \bar a_t'))
    }{\mathrm{d}\epsilon}_{|\epsilon = 0} \\
    = & \langle \sum_{t = 1}^K D_{Z_t}(P), \sum_{t = 1}^K h_{Z_t}\rangle_P, 
\end{align*}
which corresponds to the pathwise derivative of the treatment specific mean under interventions $A_t = a_t'$ and $X \sim p_X(X | \parents(X | \bar a))$ for all $X \in \bar L \cup \bar R$. This proves the rest of the EIC components $\sum_{t = 1}^K D_{Z_t}(P)$ in Section \ref{sec:EIC}. 

Lastly, note that the orthogonal decomposition gives that $\langle D_{Z_t}(P), \sum_{X \in \bar L \cup \bar R} h_X \rangle_P = 0$ and $\langle D_{R_t}(P) + D_{L_t}(P), \sum_{X \in \bar Z } h_X \rangle_P = 0$. Therefore, $        \frac{\mathrm{d}\Psi(P_\epsilon^h)}{\mathrm{d}\epsilon}_{|\epsilon = 0} = \langle \sum_{t = 1}^K D_{Z_t}(P) + D_{R_t}(P) + D_{L_t}(P),  h\rangle_P$ for all $h \in T(P)$. This proves that $D(P) = \sum_{X \in \bar L \cup \bar R \cup \bar Z} D_{X}(P)$ is the gradient and hence the unique canonical gradient in the nonparametric tangent space $T(P)$. 

\subsection{EIC of Other Longitudinal Interventions} \label{sec:CDE_EIC}

\subsubsection*{Controlled direct effect (CDE)}

Suppose that we replace the random intervention $Z_t(\bar a, \bar \Gamma^{\bar a'}) \sim \Gamma^{\bar a'}_t$ in \ref{sec:RI} with enforcing a random draw $Z_t(\bar a, \bar \Gamma) \sim \Gamma_t$ that does not depend on the control group counterfactuals. A typical choice is letting $\Gamma_t$ be a degenerated discrete density function that puts probability $1$ to a certain value $z_t$, in which case the g-computation formula for $Y(\bar a, \bar \Gamma)$ is identical to that of $Y(\bar a, \bar z)$ following static intervention. 
Another example is when we believe that some estimator $\hat\Gamma_t$ is a satisfactory proxy for the unobserved $\bar \Gamma^{\bar a'}$, and we focus on $Y(\bar a, \bar \Gamma = \bar{\hat\Gamma})$ despite the caveat that interpretation may be limited when $\bar{\hat\Gamma}$ is too deviated from $\bar \Gamma^{\bar a'}$. 
In general, these correspond to some joint intervention on $\bar A$ and $\bar Z$, where the (random/stochastic) CDE, 
$$\ex\newbracket{Y(\bar a, \bar \Gamma) - Y(\bar a', \bar \Gamma)},$$
becomes a standard average treatment effect (ATE) parameter under random intervention of both $\bar A$ and $\bar Z$. For either $\Psi(P) = \ex\newbracket{Y(\bar a, \bar \Gamma)}$ or $\ex\newbracket{Y(\bar a', \bar \Gamma)}$, $p_{Z_t}$ becomes nuisance in the sense that any submodel $P^h_{\epsilon}$ with score $h \in T_Z(P)$ leads to $\frac{\mathrm{d}\Psi(P^h_\epsilon)}{\mathrm{d}\epsilon} = 0$. This leads to $D(P) \perp T_Z(P)$ and $D_Z(P) = 0$. For $X\in\bar L \cup \bar R$, similar proof of EIC decomposition over $X\in\bar L \cup \bar R$ applies by replacing $Z_t \sim p_{Z_t}(Z_t | \parents(Z_t | \bar a'))$ with $Z_t \sim \Gamma_t(Z_t| \parents(Z_t))$ (see Lemma \ref{lemma:EIC_general}). The same methodology for NIE and NDE applies except that the EIC $D(P)$ now has $D_{Z_t} = 0$ and that a known and fixed function, $\Gamma_t(Z_t| \parents(Z_t))$, replacing $p_{Z_t}(Z_t | \parents(Z_t | \bar a'))$ in $D_{R_t}(P)$ and $D_{L_t}(P)$. 

\subsubsection*{General longitudinal stochastic intervention}

For the average treatment effect of a general longitudinal stochastic intervention (which could be a mixture of fixed and random interventions), suppose that $\bar A$ denotes the static intervention nodes and $\bar Z$ denotes the random intervention nodes onto the SCM in Section \ref{sec:RI} . At the contrast of $(\bar a, \bar\Gamma^{(1)})$ and $(\bar a', \bar\Gamma^{(2)})$ with distinct $\bar \Gamma^{(1)}$ and $\bar \Gamma^{(2)}$, one can define the generalized ATE under this mixed intervention as
$$\ex\newbracket{Y(\bar a, \bar \Gamma^{(1)}) - Y(\bar a', \bar \Gamma^{(2)})}.$$
For $\bar \Gamma = \bar \Gamma^{(1)}$ or $\bar \Gamma^{(2)}$, we have the following lemma for the EIC of $\Psi(P) = \ex\newbracket{Y(\bar a, \bar \Gamma)}$. 

\begin{lemma}\label{lemma:EIC_general}
For $\bar X = \bar R$ or $\bar L$ and $\Psi(P) = \ex\newbracket{Y(\bar a, \bar \Gamma)}$ where $\bar \Gamma$ is a set of known density function for $\bar Z$ (in the sense that $\bar Z(O)$ is a fixed function of $O$), assume that $\ex\newbracket{Y(\bar a, \bar \Gamma)}$ is identified by the similar g-computation formula as Equation \ref{eq:gcomp} but replacing $p_{Z}(Z_t | \parents(Z_t | \bar a'))$ with $\Gamma_t(Z_t | \parents(Z_t))$. Define
\begin{align*}
    Q_{R_{K + 1}}^{\bar a, \bar \Gamma} = & Y \\
    Q_{L_t}^{\bar a, \bar \Gamma}\newparethensis{\bar R_t, \bar Z_t, \bar L_{t-1}} = & \ex_{P^{\bar a, \bar \Gamma}}\newbracket{Q_{R_{t+1}}^{\bar a, \bar \Gamma} | \bar R_t, \bar Z_t, \bar L_{t-1}} = \ex_{P}\newbracket{Q_{R_{t+1}}^{\bar a, \bar \Gamma} | \bar R_t, \bar Z_t, \bar L_{t-1}, \bar A_t = \bar a_t} \\
    Q_{Z_t}^{\bar a, \bar \Gamma}\newparethensis{\bar R_t, \bar Z_{t-1}, \bar L_{t-1}} = & \ex_{P^{\bar a, \bar \Gamma}}\newbracket{Q_{L_{t}}^{\bar a, \bar \Gamma} | \bar R_t, \bar Z_{t-1}, \bar L_{t-1}} = 
    \sum_{z_t} Q_{L_{t}}^{\bar a, \bar \Gamma} \Gamma_t(z_t | \parents(Z_t)) \\
    Q_{R_t}^{\bar a, \bar \Gamma}\newparethensis{\bar R_{t-1}, \bar Z_{t-1}, \bar L_{t-1}} = & \ex_{P^{\bar a, \bar \Gamma}}\newbracket{Q_{Z_{t}}^{\bar a, \bar \Gamma} | \bar R_{t-1}, \bar Z_{t-1}, \bar L_{t-1}} = \ex_{P}\newbracket{Q_{Z_{t}}^{\bar a, \bar \Gamma} | \bar R_{t-1}, \bar Z_{t-1}, \bar L_{t-1}, \bar A_t = \bar a_t} \\
    \Psi^{\bar a, \bar \Gamma}(P) = & \ex_{P} Q_{R_1}^{\bar a, \bar \Gamma}(P)(L_0), 
\end{align*}
and
\begin{align*}
    D^{\bar a, \bar \Gamma}_{L_t} = & \frac{\indicator{\bar A_t = \bar a_t}}{\prod_{j = 1}^t p_{A}(a_j | \parents(A_j| \bar a_{j-1}))}
    \prod_{j = 1}^t \frac{\Gamma_t(Z_j | \parents(Z_j))}{p_Z(Z_j | \parents(Z_j| \bar a_j))} 
    \newbrace{Q_{R_{t+1}}^{\bar a, \bar \Gamma}(\bar R_t, \bar Z_t, \bar L_t) - 
    Q_{L_t}^{\bar a, \bar \Gamma}(\bar R_t, \bar Z_t, \bar L_{t-1})} \\
    D^{\bar a, \bar \Gamma}_{R_t} = & \frac{\indicator{\bar A_t = \bar a_t}}{\prod_{j = 1}^t p_{A}(a_j | \parents(A_j| \bar a_{j-1}))}
    \prod_{j = 1}^{t-1} \frac{\Gamma_t(Z_j | \parents(Z_j))}{p_Z(Z_j | \parents(Z_j| \bar a_j))} 
    \newbrace{Q_{Z_{t}}^{\bar a, \bar \Gamma}(\bar R_t, \bar Z_{t-1}, \bar L_{t-1}) - 
    Q_{R_{t}}^{\bar a, \bar \Gamma}(\bar R_{t-1}, \bar Z_{t-1}, \bar L_{t-1})}. 
\end{align*}    
Then the EIC of $\Psi(P)$ is $D^{\bar a, \bar \Gamma}(P) = \sum_{X_t \in \bar R \cup \bar L} D^{\bar a, \bar \Gamma}_{X_t}(P)$. 
\end{lemma}

\section{Projection Representation of EIC}\label{sec:projection_proof}

For Lemma \ref{lemma:projection}, recall that in Appendix \ref{sec:EIC_proof} we proved that $\sum_{X \in \bar R \cup \bar L} D_{X}(P)$ is the influence curve of the treatment specific mean (TSM) under interventions $A_t = a_t$ and $Z_t \sim p_{Z_t}(Z_t | \parents(Z_t | \bar a'))$. For such TSM parameter, an influence curve in the semiparametric model where $p_A$ and $p_Z$ are assumed known can be derived from the IPW estimator $\frac{1}{n}\sum_{t = 1}^K \frac{Y\indicator{\bar{A} = \bar{a}}}{\prod_{j=1}^K p_A(a_j | \parents(A_j | \bar{a}_{j-1}))} \cdot \frac{\prod_{j=1}^K p_Z(Z_j | \parents(Z_j | \bar{a}'_{j}))}{\prod_{j=1}^K p_Z(Z_j | \parents(Z_j | \bar{a}_{j}))}$ \citep{laan2018hal}. Then the influence function in the nonparametric model is the projection of this semiparametric influence curve onto the tangent space. Therefore, for $X \in \bar R \cup \bar L$, 
\begin{align*}
    D_X^{\bar a, \bar a'}(P) = \prod \newparethensis{\prod \newparethensis{G^{\bar a, \bar a'}_{LR} | L_{0}^2(P) } | T_X(P) } = \prod \newparethensis{G^{\bar a, \bar a'}_{LR} | T_X(P)}, 
\end{align*}
where $L_{0}^2(P)$ is the collection of finite variance functions $f(O)$ such that $\ex_P\newbracket{f(O)} = 0$. The same applies to $\sum_{X \in \bar Z} D_X(P)$ that is derived as the influence curve of the TSM under intervention $A_t = a_t'$, $R_t \sim p_{R_t}(R_t | \parents(R_t | \bar a))$, and $L_t \sim p_{L_t}(L_t | \parents(L_t | \bar a))$. 

Lemma \ref{lemma:projection} can also be verified with algebraic calculation using $\prod\newparethensis{f | T_X(P)} = \ex_P\newbracket{f | X, \parents(X)} - \ex_P\newbracket{f | \parents(X)}$ and expanding the integral terms of conditional expectations. 

\section{Centered HAL Basis}\label{sec:basis}

Suppose that $D_X^*(P)$ is one of the EIC component listed in Section \ref{sec:EIC} associated with the tangent space $T_X(P) = \newbrace{h(X | \parents(X)): \ex_P\newbrace{h(X | \parents(X)) | \parents(X)})=0}$. In this section, we present an alternative representation of $D_X^*(P)$ as a linear combination of centered HAL basis functions, and the TMLE updates based on the approximation. 

If we have an initial gradient $G_X(P)$ such that its projection onto the tangent space $\prod\newparethensis{G_X(P) | T_X(P)}$ equals the EIC component $D^*_X(P)$ (Lemma \ref{lemma:projection} as an example), and if $T_X(P)$ is well approximated by the linear span of a set of basis functions $\{\phi_{j, X}(P): j\}$, then we can have the following EIC representation
\begin{align*}
    D^*_X(P) = \sum_{j} \beta_{j, X}(P) \phi_{j, X}(P), 
\end{align*}
where the coefficients $\beta_X(P) = (\beta_{j, X}(P): j)$ are defined by the least squared projection
\begin{align*}
    \beta_X(P) = \arg\min_{\beta} P\newbrace{G_X(P) - \sum_{j} \beta_{j}\phi_{j, X}(P)}^2. 
\end{align*}
In practice, a large sample of size $N$ can be generated from $P$, and a follow-up cross-validated lasso regression against a subset of $\{\phi_{j, X}(P): j\}$ of size $J$ will decide at most $N-1$ non-zero coefficients in the following approximated representation with a finite sectional variation norm
\begin{align*}
    \hat{D}^*_X(P) = \sum_{j=1}^J \hat{\beta}_{j, X}(P) \phi_{j, X}(P). 
\end{align*}

Under assumption \ref{assumption_bound}, $T_X(P)$ is a subspace of the space of cadlag functions of $(X, \parents(X))$ with bounded sectional variation norms, where the latter is the space spanned by the (uncentered) HAL basis in the following form (see Section 6.2 of \cite{van2018highly}): 
\begin{align*}
    \phi_{j, X} = \indicator{U(X) \geq \mu_j}, 
\end{align*}
where $U(X) = (X, \parents(X))$. 
$u_j$ is a knot point in the range of a subvector of $U(X)$, where each element of $u_j$ takes a value in $\mathbb{R} \cup \newbrace{-\infty}$. Conducting the projection onto $T_X(P)$ in two steps gives 
\begin{align*}
        {D}^*_X(P) = \sum_{j} \beta_{j, X}(P) \phi_{j, X}(P), 
\end{align*}
where $\phi_{j, X}(P) = \Pi(\phi_{j, X} | T_X(P)) = \phi_{j, X} - \ex_P\newbracket{\phi_{j, X} | \parents(X)} =  \indicator{\parents(X) \geq \parents(x)(\mu_j)}\newparethensis{
\indicator{X \geq x(\mu_j)} - P(X \geq x(\mu_j) | \parents(X))
}. 
$
By the construction of finite-sample HAL estimator \citep{van2018highly}, using the regenerated IID sample following $P$ of size $N$, a finite subset of $\newbrace{\phi_{j, X}(P): j}$ of size $J$ can be chosen such that
the corresponding cross-validated lasso estimator $\hat{D}^*_X(P) = \sum_{j}^J \hat{\beta}_{j, X}(P) \phi_{j, X}(P)$ satisfies the convergence rate $\norm{\hat{D}^*_X(P) - D^*_X(P)}_P = O_P(N^{-1/3}(\log N)^{d/2})$.

\section{HAL-EIC TMLE}\label{sec:hal_eic_tmle_proof}

Note that there is only one additional term imposed to the expansion (\ref{eq:exact_expansion}),
\begin{align*}
    & P_n \newbrace{\pmb D^F(P_n^*) - \hat {\pmb D}^F(P_n^*)} \\
    = & (P_n - P_0)\newbrace{ {\pmb D}^F(P_n^*) - \hat {\pmb D}^F(P_n^*)} + (P_0 - P_n^*)\newbrace{ {\pmb D}^F(P_n^*) - \hat{\pmb D}^F(P_n^*)}.
\end{align*}
To maintain the asymptotic efficiency achieved under the assumptions listed in Section \ref{sec:efficiency}, 1) the first part converges under the similar Donsker condition, and 2) with the known HAL error rate \citep{bibaut2019fast} of $\norm{D^*_X(P) - \hat D^*_X(P)}_P = O_P(N^{-1/3}(\log N)^{d/2})$ where $\norm{g(O)}_P = \sqrt{\langle g, g \rangle_P} = \sqrt{Pg^2} = \sqrt{\int g(o)^2 p(o) d\mu}$ is the $L^2(P)$ norm, 
the second part (assuming the $L^2(\mu)$ norm below is applied element-wise for vectors) by the Cauchy Schwarz inequality
\begin{align*}
\norm{
(P_0 - P_n^*)\newbrace{{\pmb D}^F(P_n^*) - \hat{\pmb D}^F(P_n^*)}
}
 \leq & 
 \norm{
  \norm{p_0 - p_n^*}_\mu \cdot \norm{{\pmb D}^F(P_n^*) - \hat{\pmb D}^F(P_n^*)}_\mu 
 }, 
\end{align*}
only requires \ref{assumption_two}, that is, either $\norm{p_0 - p_n^*}_{P_0} = o_P(n^{-1/6})$ when we select $N = n$, or requires no additional condition when $N(n)$ increases at a faster rate than $n^{3/2}$, 
under the assumptions that $p_0(o) > \delta > 0$ and $p_n^*(o) > \delta > 0$ over the supports for some $\delta > 0$. 

In practice, we can simulate with $N>n$ to further improve the finite sample performance, and HAL can be included as one of the estimators in the super learner for $P_n^0$ so that $\norm{p_0 - p_n^*}_{P_0} = O_P(n^{-1/3}(\log n)^{d/2})$ is guaranteed.

\section{Multiple Robustness and Exact Remainders} \label{sec:res_proof}

First, focus on one of the dimensions $\Psi(P) = \Psi^{\bar a, \bar a'}_s(P)$ and its corresponding exact remainder as $R(P, P_0) = R^{\bar a, \bar a'}_s(P, P_0)$. Then by definition of Section \ref{sec:efficiency}, 
\begin{align*}
    R(P, P_0) & = \Psi(P) - \Psi(P_0) + P_0 D(P) \\
    & = P Q_{R_1}(P) - P_0 Q_{R_1}(P_0) + \sum_{X \in \bar R \cup \bar Z \cup \bar L} P_0 D_{X}(P) 
\end{align*}
Define the following generalized propensity terms: 
\begin{align*}
    \pi^{k}_A = & \prod_{t = 1}^k p_A(A_t | \parents(A_t)), ~ \pi^{k}_{0, A} = \prod_{t = 1}^k p_{0, A}(A_t | \parents(A_t)) \\
    \pi^{k}_R = & \prod_{t = 1}^k p_R(R_t | \parents(R_t)), ~ \pi^{k}_{0, R} = \prod_{t = 1}^k p_{0, R}(R_t | \parents(R_t)) \\
    \pi^{k}_Z = & \prod_{t = 1}^k p_Z(Z_t | \parents(Z_t)), ~ \pi^{k}_{0, Z} = \prod_{t = 1}^k p_{0, Z}(Z_t | \parents(Z_t)) \\
    \pi^{k}_L = & \prod_{t = 1}^k p_L(L_t | \parents(L_t)), ~ \pi^{k}_{0, L} = \prod_{t = 1}^k p_{0, L}(L_t | \parents(L_t)) \\
    \pi^{*k, \bar a}_A = & \prod_{t = 1}^k \indicator{A_t = a_t} \\ 
    \pi^{*k, \bar a'}_A = & \prod_{t = 1}^k \indicator{A_t = a_t'} \\
    \pi^{*k}_R = & \prod_{t = 1}^k p_R(R_t | \parents(R_t | \bar a_t)), ~ \pi^{*k}_{0, R} = \prod_{t = 1}^k p_{0, R}(R_t | \parents(R_t | \bar a_t)) \\
    \pi^{*k}_Z = & \prod_{t = 1}^k p_Z(Z_t | \parents(Z_t | \bar a_t')), ~ \pi^{*k}_{0, Z} = \prod_{t = 1}^k p_{0, Z}(Z_t | \parents(Z_t | \bar a_t')) \\
    \pi^{*k}_L = & \prod_{t = 1}^k p_L(L_t | \parents(L_t | \bar a_t)), ~ \pi^{*k}_{0, L} = \prod_{t = 1}^k p_{0, L}(L_t | \parents(L_t | \bar a_t)), 
\end{align*}
then 
\begin{align*}
    D_{L_t}(P) = & \frac{\pi^{*t, \bar a}_A \pi^{*t}_Z }{\pi^{t}_A \pi^{t}_Z}(Q_{R_{t+1}} - Q_{L_{t}} ) \\
    D_{Z_t}(P) = & \frac{\pi^{*t, \bar a'}_A \pi^{*t}_R \pi^{*,t-1}_L }{\pi^{t}_A \pi^{t}_R \pi^{t-1}_L}(Q_{L_{t}} - Q_{Z_{t}} ) \\
    D_{R_t}(P) = & \frac{\pi^{*t, \bar a}_A \pi^{*,t-1}_Z }{\pi^{t}_A \pi^{t-1}_Z}(Q_{Z_{t}} - Q_{R_{t}} ). 
\end{align*}
Plug in to the exact remainder above (let $Q_X = Q_X(P)$ when the dependence is not specified), and note that 
\begin{align*}
P_0Q_{R_1}(P_0) = & P_0 G_L(P_0) = P_0 \frac{\pi^{*K, \bar a}_A \pi^{*K}_{0, Z} }{\pi^{K}_{0, A} \pi^{K}_{0, Z}} Y \\
P_0 D_{R_1}(P) = & P_0(Q_{Z_1}(P) - Q_{R_1}(P)) \\
P_0Q_{R_1}(P_0) - P_0Q_{R_1}(P) = & P_0 \frac{\pi^{*K, \bar a}_A \pi^{*K}_{0, Z} }{\pi^{K}_{0, A} \pi^{K}_{0, Z}} (Y - Q_{L_K}(P)) + P_0 \frac{\pi^{*K, \bar a'}_A \pi^{*K}_{0, R} \pi^{*,K-1}_{0, L} }{\pi^{K}_{0, A} \pi^{K}_{0, R} \pi^{K-1}_{0, L}} (Q_{L_K}(P) - Q_{Z_K}(P)) + \cdots \\
& +  P_0(Q_{Z_1}(P) - Q_{R_1}(P)),
\end{align*}
and therefore (still let $Q_X = Q_X(P)$ for clarity)
\begin{align*}
    R(P, P_0) = & - P_0 Q_{R_1}(P_0) + P_0 Q_{R_1}(P) + \\
    & \sum_{t = 1}^K \newparethensis{
        P_0\frac{\pi^{*t, \bar a}_A \pi^{*t}_Z }{\pi^{t}_A \pi^{t}_Z}(Q_{R_{t+1}} - Q_{L_{t}} ) + 
        P_0\frac{\pi^{*t, \bar a'}_A \pi^{*t}_R \pi^{*,t-1}_L }{\pi^{t}_A \pi^{t}_R \pi^{t-1}_L}(Q_{L_{t}} - Q_{Z_{t}} ) + 
        P_0\frac{\pi^{*t, \bar a}_A \pi^{*,t-1}_Z }{\pi^{t}_A \pi^{t-1}_Z}(Q_{Z_{t}} - Q_{R_{t}} )
    } \\
    = & \sum_{t = 1}^K \Bigg(
        P_0(\frac{\pi^{*t, \bar a}_A \pi^{*t}_Z }{\pi^{t}_A \pi^{t}_Z}-\frac{\pi^{*t, \bar a}_A \pi^{*t}_{0, Z} }{\pi^{t}_{0, A} \pi^{t}_{0, Z}})(Q_{R_{t+1}} - Q_{L_{t}} ) + \\
        &  
        P_0(\frac{\pi^{*t, \bar a'}_A \pi^{*t}_R \pi^{*,t-1}_L }{\pi^{t}_A \pi^{t}_R \pi^{t-1}_L}-\frac{\pi^{*t, \bar a'}_A \pi^{*t}_{0, R} \pi^{*,t-1}_{0, L} }{\pi^{t}_{0, A} \pi^{t}_{0, R} \pi^{t-1}_{0, L}})(Q_{L_{t}} - Q_{Z_{t}} ) + \\
        &P_0(\frac{\pi^{*t, \bar a}_A \pi^{*,t-1}_{0, Z} }{\pi^{t}_A \pi^{t-1}_Z}-\frac{\pi^{*t, \bar a}_A \pi^{*,t-1}_Z }{\pi^{t}_{0, A} \pi^{t-1}_{0, Z}})(Q_{Z_{t}} - Q_{R_{t}} )
    \Bigg). 
\end{align*}
Due to the sequential definition of $\newbrace{Q_X: X \in \bar R, \bar Z, \bar L}$ as functions of $(p_R, p_Z, p_L)$, one can check that $p_{R_t} = p_{0, R_t}$ leads to $P_0\newbrace{Q_{Z_{t}} - Q_{R_{t}}} = 0$, 
$p_{Z_t} = p_{0, Z_t}$ leads to $P_0\newbrace{Q_{L_{t}} - Q_{Z_{t}}} = 0$, and 
$p_{L_t} = p_{0, L_t}$ leads to $P_0\newbrace{Q_{R_{t+1}} - Q_{L_{t}}} = 0$. 
Under positivity assumptions, this proves the statement that under one of the following three scenarios we have $R(P, P_0) = 0$: 1) $p_A = p_{0, A}$ and $p_Z = p_{0, Z}$, 2) $p_A = p_{0, A}$ and $p_R = p_{0, R}, p_L = p_{0, L}$, or 3) $p_Z = p_{0, Z}$ and $p_R = p_{0, R}, p_L = p_{0, L}$. 

Furthermore, under strong positivity and bounded variation norm assumptions as specified in \ref{assumption_bound} and \ref{assumption_two}, Cauchy-Schwarz inequality applies such that the aforementioned conditions are relaxed such that only $\norm{p_X - p_{0, X}}_P = o_P(n^{-1/4})$ is required for 1) $X \in \bar A \cup \bar Z$, 2) $X \in \bar A \cup \bar R \cup \bar L$, or 3) $X \in \bar Z \cup \bar R \cup \bar L$. This proves the multiple robustness statement in Section \ref{sec:MR}. 

\section{Numerical Results}\label{appendix:sim}

\begin{table}[ht]
\centering
\begin{tabular}{rrrrrr}
  \hline
$\ex[Y_1(\bar 1, \Gamma^{\bar 1})]$ & Bias & SD & MSE & Coverage & Width \\ 
  \hline
Exact-EIC Initial & 0.0001 & 0.0274 & 0.0008 & 0.9465 & 0.1064 \\ 
  HAL-EIC Initial & 0.0001 & 0.0274 & 0.0008 & 0.9455 & 0.1060 \\ 
  Exact-EIC TMLE & 0.0000 & 0.0274 & 0.0008 & 0.9465 & 0.1064 \\ 
  HAL-EIC TMLE & 0.0001 & 0.0274 & 0.0008 & 0.9455 & 0.1060 \\ 
   \hline
$\ex[Y_2(\bar 1, \Gamma^{\bar 1})]$ & Bias & SD & MSE & Coverage & Width \\ 
  \hline
Exact-EIC Initial & 0.0003 & 0.0282 & 0.0008 & 0.9687 & 0.1259 \\ 
  HAL-EIC Initial & 0.0004 & 0.0283 & 0.0008 & 0.9444 & 0.1085 \\ 
  Exact-EIC TMLE & 0.0005 & 0.0323 & 0.0010 & 0.9414 & 0.1251 \\ 
  HAL-EIC TMLE & 0.0004 & 0.0283 & 0.0008 & 0.9444 & 0.1085 \\ 
   \hline
$\ex[Y_1(\bar 1, \Gamma^{\bar 0})]$ & Bias & SD & MSE & Coverage & Width \\ 
  \hline
Exact-EIC Initial & 0.0001 & 0.0275 & 0.0008 & 0.9455 & 0.1086 \\ 
  HAL-EIC Initial & 0.0000 & 0.0275 & 0.0008 & 0.9434 & 0.1070 \\ 
  Exact-EIC TMLE & -0.0001 & 0.0276 & 0.0008 & 0.9455 & 0.1085 \\ 
  HAL-EIC TMLE & 0.0000 & 0.0275 & 0.0008 & 0.9434 & 0.1070 \\ 
   \hline
$\ex[Y_2(\bar 1, \Gamma^{\bar 0})]$ & Bias & SD & MSE & Coverage & Width \\ 
  \hline
Exact-EIC Initial & 0.0003 & 0.0287 & 0.0008 & 0.9737 & 0.1327 \\ 
  HAL-EIC Initial & 0.0003 & 0.0287 & 0.0008 & 0.9404 & 0.1108 \\ 
  Exact-EIC TMLE & 0.0002 & 0.0347 & 0.0012 & 0.9303 & 0.1306 \\ 
  HAL-EIC TMLE & 0.0003 & 0.0287 & 0.0008 & 0.9404 & 0.1108 \\ 
   \hline
$\ex[Y_1(\bar 0, \Gamma^{\bar 0})]$ & Bias & SD & MSE & Coverage & Width \\ 
  \hline
Exact-EIC Initial & 0.0021 & 0.0255 & 0.0007 & 0.9606 & 0.1056 \\ 
  HAL-EIC Initial & 0.0021 & 0.0255 & 0.0007 & 0.9545 & 0.1031 \\ 
  Exact-EIC TMLE & 0.0020 & 0.0261 & 0.0007 & 0.9576 & 0.1055 \\ 
  HAL-EIC TMLE & 0.0021 & 0.0255 & 0.0007 & 0.9545 & 0.1031 \\ 
   \hline
$\ex[Y_2(\bar 0, \Gamma^{\bar 0})]$ & Bias & SD & MSE & Coverage & Width \\ 
  \hline
Exact-EIC Initial & -0.0006 & 0.0267 & 0.0007 & 0.9808 & 0.1290 \\ 
  HAL-EIC Initial & -0.0006 & 0.0268 & 0.0007 & 0.9414 & 0.1051 \\ 
  Exact-EIC TMLE & -0.0005 & 0.0338 & 0.0011 & 0.9222 & 0.1266 \\ 
  HAL-EIC TMLE & -0.0006 & 0.0268 & 0.0007 & 0.9414 & 0.1051 \\ 
   \hline
\end{tabular}
\caption{Misspecification: none.} \label{tab:none}
\end{table}

\pagebreak
\begin{table}[ht]
\centering
\begin{tabular}{rrrrrr}
  \hline
$\ex[Y_1(\bar 1, \Gamma^{\bar 1})]$ & Bias & SD & MSE & Coverage & Width \\ 
  \hline
Exact-EIC Initial & 0.0002 & 0.0274 & 0.0008 & 0.8983 & 0.0894 \\ 
  HAL-EIC Initial & 0.0002 & 0.0275 & 0.0008 & 0.9003 & 0.0889 \\ 
  Exact-EIC TMLE & 0.0002 & 0.0274 & 0.0008 & 0.9003 & 0.0894 \\ 
  HAL-EIC TMLE & 0.0002 & 0.0275 & 0.0008 & 0.9003 & 0.0889 \\ 
   \hline
$\ex[Y_2(\bar 1, \Gamma^{\bar 1})]$ & Bias & SD & MSE & Coverage & Width \\ 
  \hline
Exact-EIC Initial & 0.0004 & 0.0284 & 0.0008 & 0.8983 & 0.0940 \\ 
  HAL-EIC Initial & 0.0003 & 0.0284 & 0.0008 & 0.8602 & 0.0865 \\ 
  Exact-EIC TMLE & 0.0007 & 0.0318 & 0.0010 & 0.8592 & 0.0937 \\ 
  HAL-EIC TMLE & 0.0003 & 0.0284 & 0.0008 & 0.8602 & 0.0865 \\ 
   \hline
$\ex[Y_1(\bar 1, \Gamma^{\bar 0})]$ & Bias & SD & MSE & Coverage & Width \\ 
  \hline
Exact-EIC Initial & 0.0001 & 0.0275 & 0.0008 & 0.9075 & 0.0914 \\ 
  HAL-EIC Initial & 0.0002 & 0.0276 & 0.0008 & 0.9034 & 0.0903 \\ 
  Exact-EIC TMLE & 0.0000 & 0.0277 & 0.0008 & 0.9096 & 0.0913 \\ 
  HAL-EIC TMLE & 0.0002 & 0.0276 & 0.0008 & 0.9034 & 0.0903 \\ 
   \hline
$\ex[Y_2(\bar 1, \Gamma^{\bar 0})]$ & Bias & SD & MSE & Coverage & Width \\ 
  \hline
Exact-EIC Initial & 0.0003 & 0.0289 & 0.0008 & 0.9013 & 0.0990 \\ 
  HAL-EIC Initial & 0.0003 & 0.0289 & 0.0008 & 0.8654 & 0.0890 \\ 
  Exact-EIC TMLE & 0.0004 & 0.0333 & 0.0011 & 0.8602 & 0.0983 \\ 
  HAL-EIC TMLE & 0.0003 & 0.0289 & 0.0008 & 0.8654 & 0.0890 \\ 
   \hline
$\ex[Y_1(\bar 0, \Gamma^{\bar 0})]$ & Bias & SD & MSE & Coverage & Width \\ 
  \hline
Exact-EIC Initial & 0.0021 & 0.0255 & 0.0007 & 0.9609 & 0.1048 \\ 
  HAL-EIC Initial & 0.0022 & 0.0255 & 0.0007 & 0.9579 & 0.1043 \\ 
  Exact-EIC TMLE & 0.0021 & 0.0255 & 0.0007 & 0.9568 & 0.1048 \\ 
  HAL-EIC TMLE & 0.0022 & 0.0255 & 0.0007 & 0.9579 & 0.1043 \\ 
   \hline
$\ex[Y_2(\bar 0, \Gamma^{\bar 0})]$ & Bias & SD & MSE & Coverage & Width \\ 
  \hline
Exact-EIC Initial & -0.0005 & 0.0268 & 0.0007 & 0.9712 & 0.1187 \\ 
  HAL-EIC Initial & -0.0004 & 0.0269 & 0.0007 & 0.9353 & 0.1012 \\ 
  Exact-EIC TMLE & -0.0001 & 0.0298 & 0.0009 & 0.9466 & 0.1185 \\ 
  HAL-EIC TMLE & -0.0004 & 0.0269 & 0.0007 & 0.9353 & 0.1012 \\ 
   \hline
\end{tabular}
\caption{Misspecification: A.} 
\end{table}

\pagebreak
\begin{table}[ht]
\centering
\begin{tabular}{rrrrrr}
  \hline
$\ex[Y_1(\bar 1, \Gamma^{\bar 1})]$ & Bias & SD & MSE & Coverage & Width \\ 
  \hline
Exact-EIC Initial & -0.0000 & 0.0274 & 0.0007 & 0.9477 & 0.1064 \\ 
  HAL-EIC Initial & 0.0001 & 0.0274 & 0.0008 & 0.9443 & 0.1061 \\ 
  Exact-EIC TMLE & 0.0001 & 0.0274 & 0.0007 & 0.9477 & 0.1064 \\ 
  HAL-EIC TMLE & 0.0002 & 0.0274 & 0.0008 & 0.9455 & 0.1061 \\ 
   \hline
$\ex[Y_2(\bar 1, \Gamma^{\bar 1})]$ & Bias & SD & MSE & Coverage & Width \\ 
  \hline
Exact-EIC Initial & 0.0003 & 0.0287 & 0.0008 & 0.9682 & 0.1260 \\ 
  HAL-EIC Initial & 0.0003 & 0.0287 & 0.0008 & 0.9318 & 0.1087 \\ 
  Exact-EIC TMLE & 0.0007 & 0.0329 & 0.0011 & 0.9364 & 0.1253 \\ 
  HAL-EIC TMLE & 0.0005 & 0.0287 & 0.0008 & 0.9352 & 0.1086 \\ 
   \hline
$\ex[Y_1(\bar 1, \Gamma^{\bar 0})]$ & Bias & SD & MSE & Coverage & Width \\ 
  \hline
Exact-EIC Initial & 0.0016 & 0.0274 & 0.0008 & 0.9511 & 0.1065 \\ 
  HAL-EIC Initial & 0.0017 & 0.0274 & 0.0008 & 0.9455 & 0.1062 \\ 
  Exact-EIC TMLE & 0.0002 & 0.0277 & 0.0008 & 0.9455 & 0.1068 \\ 
  HAL-EIC TMLE & 0.0008 & 0.0274 & 0.0008 & 0.9420 & 0.1062 \\ 
   \hline
$\ex[Y_2(\bar 1, \Gamma^{\bar 0})]$ & Bias & SD & MSE & Coverage & Width \\ 
  \hline
Exact-EIC Initial & 0.0017 & 0.0287 & 0.0008 & 0.9693 & 0.1261 \\ 
  HAL-EIC Initial & 0.0017 & 0.0287 & 0.0008 & 0.9341 & 0.1094 \\ 
  Exact-EIC TMLE & -0.0000 & 0.0345 & 0.0012 & 0.9273 & 0.1263 \\ 
  HAL-EIC TMLE & 0.0008 & 0.0290 & 0.0008 & 0.9284 & 0.1094 \\ 
   \hline
$\ex[Y_1(\bar 0, \Gamma^{\bar 0})]$ & Bias & SD & MSE & Coverage & Width \\ 
  \hline
Exact-EIC Initial & 0.0041 & 0.0257 & 0.0007 & 0.9568 & 0.1055 \\ 
  HAL-EIC Initial & 0.0041 & 0.0258 & 0.0007 & 0.9534 & 0.1031 \\ 
  Exact-EIC TMLE & 0.0029 & 0.0261 & 0.0007 & 0.9580 & 0.1055 \\ 
  HAL-EIC TMLE & 0.0031 & 0.0255 & 0.0007 & 0.9591 & 0.1031 \\ 
   \hline
$\ex[Y_2(\bar 0, \Gamma^{\bar 0})]$ & Bias & SD & MSE & Coverage & Width \\ 
  \hline
Exact-EIC Initial & 0.0016 & 0.0277 & 0.0008 & 0.9773 & 0.1291 \\ 
  HAL-EIC Initial & 0.0016 & 0.0278 & 0.0008 & 0.9352 & 0.1048 \\ 
  Exact-EIC TMLE & 0.0006 & 0.0342 & 0.0012 & 0.9136 & 0.1268 \\ 
  HAL-EIC TMLE & 0.0006 & 0.0273 & 0.0007 & 0.9455 & 0.1048 \\ 
   \hline
\end{tabular}
\caption{Misspecification: Z.} 
\end{table}

\pagebreak
\begin{table}[ht]
\centering
\begin{tabular}{rrrrrr}
  \hline
$\ex[Y_1(\bar 1, \Gamma^{\bar 1})]$ & Bias & SD & MSE & Coverage & Width \\ 
  \hline
Exact-EIC Initial & 0.0532 & 0.0186 & 0.0032 & 0.5111 & 0.1072 \\ 
  HAL-EIC Initial & 0.0533 & 0.0188 & 0.0032 & 0.5069 & 0.1070 \\ 
  Exact-EIC TMLE & 0.0003 & 0.0273 & 0.0007 & 0.9504 & 0.1068 \\ 
  HAL-EIC TMLE & 0.0001 & 0.0276 & 0.0008 & 0.9409 & 0.1063 \\ 
   \hline
$\ex[Y_2(\bar 1, \Gamma^{\bar 1})]$ & Bias & SD & MSE & Coverage & Width \\ 
  \hline
Exact-EIC Initial & 0.0494 & 0.0183 & 0.0028 & 0.8110 & 0.1281 \\ 
  HAL-EIC Initial & 0.0494 & 0.0183 & 0.0028 & 0.6473 & 0.1105 \\ 
  Exact-EIC TMLE & 0.0005 & 0.0320 & 0.0010 & 0.9525 & 0.1263 \\ 
  HAL-EIC TMLE & -0.0005 & 0.0279 & 0.0008 & 0.9472 & 0.1118 \\ 
   \hline
$\ex[Y_1(\bar 1, \Gamma^{\bar 0})]$ & Bias & SD & MSE & Coverage & Width \\ 
  \hline
Exact-EIC Initial & 0.0549 & 0.0186 & 0.0034 & 0.4984 & 0.1096 \\ 
  HAL-EIC Initial & 0.0549 & 0.0187 & 0.0034 & 0.4794 & 0.1082 \\ 
  Exact-EIC TMLE & 0.0005 & 0.0276 & 0.0008 & 0.9493 & 0.1090 \\ 
  HAL-EIC TMLE & 0.0009 & 0.0278 & 0.0008 & 0.9440 & 0.1074 \\ 
   \hline
$\ex[Y_2(\bar 1, \Gamma^{\bar 0})]$ & Bias & SD & MSE & Coverage & Width \\ 
  \hline
Exact-EIC Initial & 0.0508 & 0.0183 & 0.0029 & 0.8332 & 0.1354 \\ 
  HAL-EIC Initial & 0.0508 & 0.0183 & 0.0029 & 0.6547 & 0.1134 \\ 
  Exact-EIC TMLE & 0.0002 & 0.0342 & 0.0012 & 0.9356 & 0.1319 \\ 
  HAL-EIC TMLE & -0.0001 & 0.0283 & 0.0008 & 0.9535 & 0.1147 \\ 
   \hline
$\ex[Y_1(\bar 0, \Gamma^{\bar 0})]$ & Bias & SD & MSE & Coverage & Width \\ 
  \hline
Exact-EIC Initial & 0.0590 & 0.0186 & 0.0038 & 0.3749 & 0.1067 \\ 
  HAL-EIC Initial & 0.0590 & 0.0188 & 0.0038 & 0.3601 & 0.1045 \\ 
  Exact-EIC TMLE & 0.0021 & 0.0260 & 0.0007 & 0.9599 & 0.1057 \\ 
  HAL-EIC TMLE & 0.0020 & 0.0255 & 0.0007 & 0.9588 & 0.1037 \\ 
   \hline
$\ex[Y_2(\bar 0, \Gamma^{\bar 0})]$ & Bias & SD & MSE & Coverage & Width \\ 
  \hline
Exact-EIC Initial & 0.0536 & 0.0183 & 0.0032 & 0.7635 & 0.1330 \\ 
  HAL-EIC Initial & 0.0537 & 0.0184 & 0.0032 & 0.5586 & 0.1105 \\ 
  Exact-EIC TMLE & -0.0005 & 0.0342 & 0.0012 & 0.9166 & 0.1277 \\ 
  HAL-EIC TMLE & -0.0002 & 0.0271 & 0.0007 & 0.9599 & 0.1119 \\ 
   \hline
\end{tabular}
\caption{Misspecification: Y.} \label{tab:Y}
\end{table}

\clearpage

\begin{table}[ht]
\centering
\begin{tabular}{rrrrrr}
  \hline
$\ex[Y_1(\bar 1, \Gamma^{\bar 1})]$ & Bias & SD & MSE & Coverage & Width \\ 
  \hline
Exact-EIC Initial & -0.0000 & 0.0251 & 0.0006 & 0.9394 & 0.0988 \\ 
  HAL-EIC Initial & -0.0001 & 0.0251 & 0.0006 & 0.9414 & 0.0978 \\ 
  Exact-EIC TMLE & -0.0001 & 0.0252 & 0.0006 & 0.9455 & 0.0988 \\ 
  HAL-EIC TMLE & -0.0001 & 0.0251 & 0.0006 & 0.9414 & 0.0978 \\ 
   \hline
$\ex[Y_2(\bar 1, \Gamma^{\bar 1})]$ & Bias & SD & MSE & Coverage & Width \\ 
  \hline
Exact-EIC Initial & -0.0001 & 0.0254 & 0.0006 & 0.9697 & 0.1142 \\ 
  HAL-EIC Initial & -0.0002 & 0.0254 & 0.0006 & 0.9404 & 0.0988 \\ 
  Exact-EIC TMLE & 0.0004 & 0.0301 & 0.0009 & 0.9273 & 0.1129 \\ 
  HAL-EIC TMLE & -0.0002 & 0.0254 & 0.0006 & 0.9404 & 0.0988 \\ 
   \hline
$\ex[Y_1(\bar 1, \Gamma^{\bar 0})]$ & Bias & SD & MSE & Coverage & Width \\ 
  \hline
Exact-EIC Initial & 0.0001 & 0.0254 & 0.0006 & 0.9424 & 0.1010 \\ 
  HAL-EIC Initial & 0.0001 & 0.0254 & 0.0006 & 0.9424 & 0.0990 \\ 
  Exact-EIC TMLE & -0.0000 & 0.0256 & 0.0007 & 0.9414 & 0.1009 \\ 
  HAL-EIC TMLE & 0.0001 & 0.0254 & 0.0006 & 0.9424 & 0.0990 \\ 
   \hline
$\ex[Y_2(\bar 1, \Gamma^{\bar 0})]$ & Bias & SD & MSE & Coverage & Width \\ 
  \hline
Exact-EIC Initial & -0.0000 & 0.0260 & 0.0007 & 0.9768 & 0.1209 \\ 
  HAL-EIC Initial & -0.0000 & 0.0260 & 0.0007 & 0.9485 & 0.1036 \\ 
  Exact-EIC TMLE & 0.0006 & 0.0330 & 0.0011 & 0.9253 & 0.1185 \\ 
  HAL-EIC TMLE & -0.0000 & 0.0260 & 0.0007 & 0.9485 & 0.1036 \\ 
   \hline
$\ex[Y_1(\bar 0, \Gamma^{\bar 0})]$ & Bias & SD & MSE & Coverage & Width \\ 
  \hline
Exact-EIC Initial & 0.0032 & 0.0296 & 0.0009 & 0.9677 & 0.1274 \\ 
  HAL-EIC Initial & 0.0032 & 0.0296 & 0.0009 & 0.9515 & 0.1171 \\ 
  Exact-EIC TMLE & 0.0035 & 0.0321 & 0.0010 & 0.9475 & 0.1270 \\ 
  HAL-EIC TMLE & 0.0032 & 0.0296 & 0.0009 & 0.9515 & 0.1171 \\ 
   \hline
$\ex[Y_2(\bar 0, \Gamma^{\bar 0})]$ & Bias & SD & MSE & Coverage & Width \\ 
  \hline
Exact-EIC Initial & 0.0007 & 0.0312 & 0.0010 & 0.9869 & 0.1896 \\ 
  HAL-EIC Initial & 0.0008 & 0.0312 & 0.0010 & 0.9323 & 0.1227 \\ 
  Exact-EIC TMLE & 0.0004 & 0.0528 & 0.0028 & 0.7869 & 0.1613 \\ 
  HAL-EIC TMLE & 0.0008 & 0.0312 & 0.0010 & 0.9323 & 0.1227 \\ 
   \hline
\end{tabular}
\caption{Misspecification: none, positivity scaling factor $\lambda = 2$.} 
\end{table}

\pagebreak
\begin{table}[ht]
\centering
\begin{tabular}{rrrrrr}
  \hline
$\ex[Y_1(\bar 1, \Gamma^{\bar 1})]$ & Bias & SD & MSE & Coverage & Width \\ 
  \hline
Exact-EIC Initial & 0.0010 & 0.0239 & 0.0006 & 0.9391 & 0.0948 \\ 
  HAL-EIC Initial & 0.0012 & 0.0240 & 0.0006 & 0.9359 & 0.0930 \\ 
  Exact-EIC TMLE & 0.0008 & 0.0243 & 0.0006 & 0.9391 & 0.0948 \\ 
  HAL-EIC TMLE & 0.0012 & 0.0240 & 0.0006 & 0.9359 & 0.0930 \\ 
   \hline
$\ex[Y_2(\bar 1, \Gamma^{\bar 1})]$ & Bias & SD & MSE & Coverage & Width \\ 
  \hline
Exact-EIC Initial & 0.0012 & 0.0235 & 0.0006 & 0.9695 & 0.1100 \\ 
  HAL-EIC Initial & 0.0013 & 0.0235 & 0.0006 & 0.9475 & 0.0931 \\ 
  Exact-EIC TMLE & 0.0015 & 0.0289 & 0.0008 & 0.9233 & 0.1077 \\ 
  HAL-EIC TMLE & 0.0013 & 0.0235 & 0.0006 & 0.9475 & 0.0931 \\ 
   \hline
$\ex[Y_1(\bar 1, \Gamma^{\bar 0})]$ & Bias & SD & MSE & Coverage & Width \\ 
  \hline
Exact-EIC Initial & 0.0011 & 0.0242 & 0.0006 & 0.9370 & 0.0970 \\ 
  HAL-EIC Initial & 0.0013 & 0.0243 & 0.0006 & 0.9296 & 0.0947 \\ 
  Exact-EIC TMLE & 0.0007 & 0.0247 & 0.0006 & 0.9370 & 0.0971 \\ 
  HAL-EIC TMLE & 0.0013 & 0.0243 & 0.0006 & 0.9296 & 0.0947 \\ 
   \hline
$\ex[Y_2(\bar 1, \Gamma^{\bar 0})]$ & Bias & SD & MSE & Coverage & Width \\ 
  \hline
Exact-EIC Initial & 0.0011 & 0.0239 & 0.0006 & 0.9737 & 0.1183 \\ 
  HAL-EIC Initial & 0.0011 & 0.0240 & 0.0006 & 0.9559 & 0.1049 \\ 
  Exact-EIC TMLE & -0.0008 & 0.0346 & 0.0012 & 0.8897 & 0.1172 \\ 
  HAL-EIC TMLE & 0.0011 & 0.0240 & 0.0006 & 0.9559 & 0.1049 \\ 
   \hline
$\ex[Y_1(\bar 0, \Gamma^{\bar 0})]$ & Bias & SD & MSE & Coverage & Width \\ 
  \hline
Exact-EIC Initial & 0.0023 & 0.0343 & 0.0012 & 0.9800 & 0.1650 \\ 
  HAL-EIC Initial & 0.0023 & 0.0344 & 0.0012 & 0.9443 & 0.1334 \\ 
  Exact-EIC TMLE & 0.0010 & 0.0430 & 0.0019 & 0.9349 & 0.1624 \\ 
  HAL-EIC TMLE & 0.0023 & 0.0344 & 0.0012 & 0.9443 & 0.1334 \\ 
   \hline
$\ex[Y_2(\bar 0, \Gamma^{\bar 0})]$ & Bias & SD & MSE & Coverage & Width \\ 
  \hline
Exact-EIC Initial & 0.0004 & 0.0354 & 0.0013 & 0.9706 & 0.2595 \\ 
  HAL-EIC Initial & 0.0004 & 0.0354 & 0.0013 & 0.8845 & 0.1411 \\ 
  Exact-EIC TMLE & 0.0028 & 0.0802 & 0.0064 & 0.6113 & 0.1732 \\ 
  HAL-EIC TMLE & 0.0004 & 0.0354 & 0.0013 & 0.8845 & 0.1411 \\ 
   \hline
\end{tabular}
\caption{Misspecification: none, positivity scaling factor $\lambda = 3$.} 
\end{table}

\pagebreak
\begin{table}[ht]
\centering
\begin{tabular}{rrrrrr}
  \hline
$\ex[Y_1(\bar 1, \Gamma^{\bar 1})]$ & Bias & SD & MSE & Coverage & Width \\ 
  \hline
Exact-EIC Initial & 0.0009 & 0.0230 & 0.0005 & 0.9619 & 0.0927 \\ 
  HAL-EIC Initial & 0.0009 & 0.0231 & 0.0005 & 0.9499 & 0.0902 \\ 
  Exact-EIC TMLE & 0.0008 & 0.0236 & 0.0006 & 0.9459 & 0.0927 \\ 
  HAL-EIC TMLE & 0.0009 & 0.0231 & 0.0005 & 0.9499 & 0.0902 \\ 
   \hline
$\ex[Y_2(\bar 1, \Gamma^{\bar 1})]$ & Bias & SD & MSE & Coverage & Width \\ 
  \hline
Exact-EIC Initial & 0.0008 & 0.0227 & 0.0005 & 0.9709 & 0.1094 \\ 
  HAL-EIC Initial & 0.0008 & 0.0228 & 0.0005 & 0.9379 & 0.0894 \\ 
  Exact-EIC TMLE & 0.0008 & 0.0294 & 0.0009 & 0.8998 & 0.1054 \\ 
  HAL-EIC TMLE & 0.0008 & 0.0228 & 0.0005 & 0.9379 & 0.0894 \\ 
   \hline
$\ex[Y_1(\bar 1, \Gamma^{\bar 0})]$ & Bias & SD & MSE & Coverage & Width \\ 
  \hline
Exact-EIC Initial & 0.0009 & 0.0234 & 0.0005 & 0.9619 & 0.0952 \\ 
  HAL-EIC Initial & 0.0009 & 0.0234 & 0.0005 & 0.9519 & 0.0928 \\ 
  Exact-EIC TMLE & 0.0005 & 0.0249 & 0.0006 & 0.9439 & 0.0963 \\ 
  HAL-EIC TMLE & 0.0009 & 0.0234 & 0.0005 & 0.9519 & 0.0928 \\ 
   \hline
$\ex[Y_2(\bar 1, \Gamma^{\bar 0})]$ & Bias & SD & MSE & Coverage & Width \\ 
  \hline
Exact-EIC Initial & 0.0007 & 0.0232 & 0.0005 & 0.9749 & 0.1221 \\ 
  HAL-EIC Initial & 0.0006 & 0.0233 & 0.0005 & 0.9449 & 0.1082 \\ 
  Exact-EIC TMLE & -0.0043 & 0.0392 & 0.0016 & 0.8347 & 0.1473 \\ 
  HAL-EIC TMLE & 0.0006 & 0.0233 & 0.0005 & 0.9449 & 0.1082 \\ 
   \hline
$\ex[Y_1(\bar 0, \Gamma^{\bar 0})]$ & Bias & SD & MSE & Coverage & Width \\ 
  \hline
Exact-EIC Initial & 0.0025 & 0.0369 & 0.0014 & 0.9950 & 0.2265 \\ 
  HAL-EIC Initial & 0.0025 & 0.0369 & 0.0014 & 0.9629 & 0.1506 \\ 
  Exact-EIC TMLE & 0.0028 & 0.0605 & 0.0037 & 0.8707 & 0.2123 \\ 
  HAL-EIC TMLE & 0.0025 & 0.0369 & 0.0014 & 0.9629 & 0.1506 \\ 
   \hline
$\ex[Y_2(\bar 0, \Gamma^{\bar 0})]$ & Bias & SD & MSE & Coverage & Width \\ 
  \hline
Exact-EIC Initial & 0.0021 & 0.0393 & 0.0015 & 0.9509 & 0.3478 \\ 
  HAL-EIC Initial & 0.0020 & 0.0393 & 0.0015 & 0.8307 & 0.1562 \\ 
  Exact-EIC TMLE & 0.0105 & 0.1101 & 0.0122 & 0.4559 & 0.2778 \\ 
  HAL-EIC TMLE & 0.0020 & 0.0393 & 0.0015 & 0.8307 & 0.1562 \\ 
   \hline
\end{tabular}
\caption{Misspecification: none, positivity scaling factor $\lambda = 4$.} 
\end{table}

\pagebreak
\begin{table}[ht]
\centering
\begin{tabular}{rrrrrr}
  \hline
$\ex[Y_1(\bar 1, \Gamma^{\bar 1})]$ & Bias & SD & MSE & Coverage & Width \\ 
  \hline
Exact-EIC Initial & 0.0013 & 0.0226 & 0.0005 & 0.9604 & 0.0919 \\ 
  HAL-EIC Initial & 0.0013 & 0.0228 & 0.0005 & 0.9451 & 0.0886 \\ 
  Exact-EIC TMLE & 0.0010 & 0.0234 & 0.0005 & 0.9502 & 0.0918 \\ 
  HAL-EIC TMLE & 0.0013 & 0.0228 & 0.0005 & 0.9451 & 0.0886 \\ 
   \hline
$\ex[Y_2(\bar 1, \Gamma^{\bar 1})]$ & Bias & SD & MSE & Coverage & Width \\ 
  \hline
Exact-EIC Initial & 0.0010 & 0.0222 & 0.0005 & 0.9787 & 0.1111 \\ 
  HAL-EIC Initial & 0.0010 & 0.0223 & 0.0005 & 0.9360 & 0.0871 \\ 
  Exact-EIC TMLE & 0.0010 & 0.0303 & 0.0009 & 0.8770 & 0.1054 \\ 
  HAL-EIC TMLE & 0.0010 & 0.0223 & 0.0005 & 0.9360 & 0.0871 \\ 
   \hline
$\ex[Y_1(\bar 1, \Gamma^{\bar 0})]$ & Bias & SD & MSE & Coverage & Width \\ 
  \hline
Exact-EIC Initial & 0.0014 & 0.0229 & 0.0005 & 0.9644 & 0.0945 \\ 
  HAL-EIC Initial & 0.0013 & 0.0231 & 0.0005 & 0.9543 & 0.0925 \\ 
  Exact-EIC TMLE & 0.0004 & 0.0273 & 0.0007 & 0.9360 & 0.0970 \\ 
  HAL-EIC TMLE & 0.0013 & 0.0231 & 0.0005 & 0.9543 & 0.0925 \\ 
   \hline
$\ex[Y_2(\bar 1, \Gamma^{\bar 0})]$ & Bias & SD & MSE & Coverage & Width \\ 
  \hline
Exact-EIC Initial & 0.0009 & 0.0226 & 0.0005 & 0.9817 & 0.1211 \\ 
  HAL-EIC Initial & 0.0009 & 0.0227 & 0.0005 & 0.9502 & 0.1141 \\ 
  Exact-EIC TMLE & -0.0089 & 0.0451 & 0.0021 & 0.7764 & 0.1370 \\ 
  HAL-EIC TMLE & 0.0009 & 0.0227 & 0.0005 & 0.9502 & 0.1141 \\ 
   \hline
$\ex[Y_1(\bar 0, \Gamma^{\bar 0})]$ & Bias & SD & MSE & Coverage & Width \\ 
  \hline
Exact-EIC Initial & 0.0016 & 0.0413 & 0.0017 & 0.9919 & 0.2520 \\ 
  HAL-EIC Initial & 0.0015 & 0.0414 & 0.0017 & 0.9512 & 0.1623 \\ 
  Exact-EIC TMLE & 0.0017 & 0.0897 & 0.0080 & 0.7083 & 0.2234 \\ 
  HAL-EIC TMLE & 0.0015 & 0.0414 & 0.0017 & 0.9512 & 0.1623 \\ 
   \hline
$\ex[Y_2(\bar 0, \Gamma^{\bar 0})]$ & Bias & SD & MSE & Coverage & Width \\ 
  \hline
Exact-EIC Initial & 0.0019 & 0.0421 & 0.0018 & 0.9350 & 0.3276 \\ 
  HAL-EIC Initial & 0.0019 & 0.0421 & 0.0018 & 0.7683 & 0.1672 \\ 
  Exact-EIC TMLE & 0.0077 & 0.1348 & 0.0182 & 0.3780 & 0.2262 \\ 
  HAL-EIC TMLE & 0.0019 & 0.0421 & 0.0018 & 0.7683 & 0.1672 \\ 
   \hline
\end{tabular}
\caption{Misspecification: none, positivity scaling factor $\lambda = 5$.} 
\end{table}

\pagebreak

\clearpage

\section{List of notations}\label{appendix_notations}

\pmb{Data and Model}

\begin{tabular}{cp{0.9\textwidth}}
$\bar X$ & the available history of $X$ such as $\newparethensis{X_t: t = 0, \dots K}$ \\
$\bar X_t$ & the available history of $X$ till $t$ such as $\newparethensis{X_s: s = 0, \dots t}$ \\
$\bar X_s^t$ & the available history of $X$ from $s$ to $t$ (assuming $s\leq t$)\\
$x$ & a realization value in the range of a random variable $X$ \\
$\parents(X)$ & the parent nodes prior to $X$ given a variable ordering \\
$\children(X)$ & the child nodes after $X$ given a variable ordering \\
$\parents(X | \bar a)$ & the vector of parent nodes of $X$ but intervening $\parents(X) \cap \bar A$ following $\bar A = \bar a$; for example, $\parents(R_1 | \bar a) = \newparethensis{L_0, a_1}$ in Section \ref{sec:data}\\
  $\mathcal{M}$ & statistical model, a collection of data distributions\\
  $P$ & data distribution\\
  $P_n$ & empirical distribution of an IID sample $O_1, \dots, O_n$\\
  $Pf$ & expectation $Pf = \mathbb{E}_P(f) = \int f(o) dP(o)$ of $f(O)$ under $P$; for example, $P_nf = \frac{1}{n}\sum_{i=1}^n f(O_i)$\\
  $p$ & density $p = dP/d\mu$ of $P$ with respect to some measure $\mu$\\
  $p_X(X | \parents(X))$ & conditional density of variable $X$ given $\parents(X)$\\
  $P_0$ & true data distribution\\
  $X(\bar a)$ & the counterfactual of $X$ under intervention $\bar A = \bar a$ given a structural causal model \\
  $\Gamma^{\bar a'}_t$ & conditional density of $Z_t(\bar a')$ given $\bar R_t(\bar a'), \bar Z_{t-1}(\bar a'), \bar L_{t-1}(\bar a')$\\
  $X(\bar a, \bar \Gamma^{\bar a'})$ & the counterfactual of $X$ under intervention $\bar A = \bar a$ and $Z_t \sim  \Gamma^{\bar a'}_t(z_t | \bar R_t(\bar a, \Gamma^{\bar a'}), \bar Z_{t-1}(\bar a, \Gamma^{\bar a'}), \bar L_{t-1}(\bar a, \Gamma^{\bar a'}) ) $ for $t = 1, \dots, K$;  \\
 \end{tabular}
 
 \pmb{Statistical Estimation}

\begin{tabular}{cp{0.9\textwidth}}
$\Psi$ & a general $k$ dimensional target parameter $\mathcal{M} \to \mathbb{R}^k$ \\
$\Psi_s^{\bar a, \bar a'}(P)$ & $\ex\newbracket{Y_s(\bar a, \bar \Gamma^{\bar a'})}$ where $Y_s = \psi(\bar L_K)$ is the $s$-th outcome; $\psi$ is a function of $\bar L_K$ \\
$P^{\bar a, \bar a'}$ & the counterfactual of distribution $P$ under intervention $(\bar a, \bar \Gamma^{\bar a'})$ \\
$Q^{\bar a, \bar a'}_{s, X}(\parents(X) \setminus \bar A)$ & $\ex_{P^{\bar a, \bar a'}}\newbracket{Y_s | \parents(X) \setminus \bar A}$, the conditional mean (following $P^{\bar a, \bar a'}$) of $Y_s$ given the parent nodes of $X$ excluding $\bar A$\\ 
$D^{\bar a, \bar a'}_{s}(P)$ & the canonical gradient of $\Psi_s^{\bar a, \bar a'}(P)$ \\
$D^{\bar a, \bar a'}_{s, X}(P)$ & $\prod\newparethensis{D^{\bar a, \bar a'}_{s}(P) | T_X(P)}$ the gradient component of $X$ defined as the projection onto the tangent subspace $T_X(P) = \newbrace{f \in L^2(P): \ex\newbracket{f | \parents(X)} = 0}$ \\
$\tilde p_X(p_X, \bar \epsilon_{X})$ & a (multivariate) locally least favorable path through $p_X$\\
  $P_n^0$ & an initial distribution estimator for $P_0$\\
  $\tilde P_n$ & a TMLE update of $P_n^0$ \\
  $P_n^*$ & the final TMLE update\\
 \end{tabular}

\bibliographystyle{imsart-nameyear} 
\bibliography{main}

\end{document}